\documentclass[twocolumn,twocolappendix]{aastex631}
\usepackage{natbib}
\usepackage{url}
\usepackage{color}
\usepackage{hyperref}
\usepackage{amsmath}
\definecolor{red}{rgb}{1,0,0}
\definecolor{orange}{rgb}{1,0.5,0}
\definecolor{green}{rgb}{0,127,0}
\definecolor{grey}{rgb}{0.6627,0.6627,0.6627}
\definecolor{skyblue}{rgb}{0.53,0.808,0.98}

\newcommand{\lya}{Ly$\alpha$}

\newcommand{\se}{\texttt{\textsc{Source Extractor}}}

\newcommand{\eazy}{\texttt{\textsc{EAZY}}}
\newcommand{\mirage}{\texttt{\textsc{Mirage}}}
\newcommand{\pipeline}{\texttt{\textsc{jwst}}}
\newcommand{\photutils}{\texttt{\textsc{Photutils}}}

\newcommand{\hst}{\textit{HST}}
\newcommand{\spitzer}{\textit{Spitzer}}

\newcommand{\jwst}{\textit{JWST}}

\newcommand{\sersic}{S\'{e}rsic}

\newcommand\myvdots{\vbox{\baselineskip=1pt \lineskiplimit=0pt \kern10pt \hbox{.}\hbox{.}\hbox{.}}} 

\shorttitle{CEERS Epoch 1 NIRCam Imaging}
\shortauthors{Bagley et al.}

\begin{document}

\title{CEERS Epoch 1 NIRCam Imaging: Reduction Methods and Simulations Enabling Early JWST Science Results}

\correspondingauthor{Micaela B. Bagley}
\email{mbagley@utexas.edu}

\author[0000-0002-9921-9218]{Micaela B. Bagley}
\affiliation{Department of Astronomy, The University of Texas at Austin, Austin, TX, USA}

\author[0000-0001-8519-1130]{Steven L. Finkelstein}
\affiliation{Department of Astronomy, The University of Texas at Austin, Austin, TX, USA}

\author[0000-0002-6610-2048]{Anton M. Koekemoer}
\affiliation{Space Telescope Science Institute, 3700 San Martin Drive, Baltimore, MD 21218, USA}

\author[0000-0001-7113-2738]{Henry C. Ferguson}
\affiliation{Space Telescope Science Institute, 3700 San Martin Drive, Baltimore, MD 21218, USA}

\author[0000-0002-7959-8783]{Pablo Arrabal Haro}
\affiliation{NSF's National Optical-Infrared Astronomy Research Laboratory, 950 N. Cherry Ave., Tucson, AZ 85719, USA}

\author[0000-0001-5414-5131]{Mark Dickinson}
\affiliation{NSF's National Optical-Infrared Astronomy Research Laboratory, 950 N. Cherry Ave., Tucson, AZ 85719, USA}

\author[0000-0001-9187-3605]{Jeyhan S. Kartaltepe}
\affiliation{Laboratory for Multiwavelength Astrophysics, School of Physics and Astronomy, Rochester Institute of Technology, 84 Lomb Memorial Drive, Rochester, NY 14623, USA}

\author[0000-0001-7503-8482]{Casey Papovich}
\affiliation{Department of Physics and Astronomy, Texas A\&M University, College Station, TX, 77843-4242 USA}
\affiliation{George P.\ and Cynthia Woods Mitchell Institute for Fundamental Physics and Astronomy, Texas A\&M University, College Station, TX, 77843-4242 USA}

\author[0000-0003-4528-5639]{Pablo G. P\'erez-Gonz\'alez}
\affiliation{Centro de Astrobiolog\'{\i}a (CAB), CSIC-INTA, Ctra. de Ajalvir km 4, Torrej\'on de Ardoz, E-28850, Madrid, Spain}

\author[0000-0003-3382-5941]{Nor Pirzkal}
\affiliation{ESA/AURA Space Telescope Science Institute}

\author[0000-0002-6748-6821]{Rachel S. Somerville}
\affiliation{Center for Computational Astrophysics, Flatiron Institute, 162 5th Avenue, New York, NY, 10010, USA}

\author[0000-0001-9262-9997]{Christopher N. A. Willmer}
\affiliation{Steward Observatory, University of Arizona, 933 N. Cherry Ave., Tucson, AZ, 85721, USA}

\author[0000-0001-8835-7722]{Guang Yang}
\affiliation{Kapteyn Astronomical Institute, University of Groningen, P.O. Box 800, 9700 AV Groningen, The Netherlands}
\affiliation{SRON Netherlands Institute for Space Research, Postbus 800, 9700 AV Groningen, The Netherlands}

\author[0000-0003-3466-035X]{L. Y. Aaron\ Yung}
\altaffiliation{NASA Postdoctoral Fellow}
\affiliation{Astrophysics Science Division, NASA Goddard Space Flight Center, 8800 Greenbelt Rd, Greenbelt, MD 20771, USA}

\author[0000-0003-3820-2823]{Adriano Fontana}
\affiliation{INAF - Osservatorio Astronomico di Roma, via di Frascati 33, 00078 Monte Porzio Catone, Italy}

\author[0000-0002-5688-0663]{Andrea Grazian}
\affiliation{INAF--Osservatorio Astronomico di Padova, Vicolo dell'Osservatorio 5, I-35122, Padova, Italy}

\author[0000-0001-9440-8872]{Norman A. Grogin}
\affiliation{Space Telescope Science Institute, 3700 San Martin Drive, Baltimore, MD 21218, USA}

\author[0000-0002-3301-3321]{Michaela Hirschmann}
\affiliation{Institute of Physics, Laboratory of Galaxy Evolution, Ecole Polytechnique Fédérale de Lausanne (EPFL), Observatoire de Sauverny, 1290 Versoix, Switzerland}

\author[0000-0001-8152-3943]{Lisa J. Kewley}
\affiliation{Center for Astrophysics | Harvard \& Smithsonian, 60 Garden Street, Cambridge, MA 02138, USA}

\author[0000-0002-5537-8110]{Allison Kirkpatrick}
\affiliation{Department of Physics and Astronomy, University of Kansas, Lawrence, KS 66045, USA}

\author[0000-0002-8360-3880]{Dale D. Kocevski}
\affiliation{Department of Physics and Astronomy, Colby College, Waterville, ME 04901, USA}

\author[0000-0003-3130-5643]{Jennifer M. Lotz}
\affiliation{Gemini Observatory/NSF's National Optical-Infrared Astronomy Research Laboratory, 950 N. Cherry Ave., Tucson, AZ 85719, USA}

\author[0000-0002-8450-9992]{Aubrey Medrano}
\affiliation{University of Massachusetts Amherst, 710 North Pleasant Street, Amherst, MA 01003-9305, USA}

\author[0000-0003-4965-0402]{Alexa M.\ Morales}
\affiliation{Department of Astronomy, The University of Texas at Austin, Austin, TX, USA}

\author[0000-0001-8940-6768]{Laura Pentericci}
\affiliation{INAF - Osservatorio Astronomico di Roma, via di Frascati 33, 00078 Monte Porzio Catone, Italy}

\author[0000-0002-5269-6527]{Swara Ravindranath}
\affiliation{Space Telescope Science Institute, 3700 San Martin Drive, Baltimore, MD 21218, USA}

\author[0000-0002-1410-0470]{Jonathan R. Trump}
\affiliation{Department of Physics, 196 Auditorium Road, Unit 3046, University of Connecticut, Storrs, CT 06269, USA}

\author[0000-0003-3903-6935]{Stephen M.~Wilkins} %
\affiliation{Astronomy Centre, University of Sussex, Falmer, Brighton BN1 9QH, UK}
\affiliation{Institute of Space Sciences and Astronomy, University of Malta, Msida MSD 2080, Malta}

\author[0000-0003-2536-1614]{Antonello Calabr{\`o}} 
\affiliation{INAF - Osservatorio Astronomico di Roma, via di Frascati 33, 00078 Monte Porzio Catone, Italy}

\author[0000-0003-1371-6019]{M. C. Cooper}
\affiliation{Department of Physics \& Astronomy, University of California, Irvine, 4129 Reines Hall, Irvine, CA 92697, USA}

\author[0000-0001-6820-0015]{Luca Costantin}
\affiliation{Centro de Astrobiolog\'ia (CSIC-INTA), Ctra de Ajalvir km 4, Torrej\'on de Ardoz, 28850, Madrid, Spain}

\author[0000-0002-6219-5558]{Alexander de la Vega}
\affiliation{Department of Physics and Astronomy, University of California, 900 University Ave, Riverside, CA 92521, USA}

\author[0000-0001-6251-4988]{Taylor A. Hutchison}
\altaffiliation{NASA Postdoctoral Fellow}
\affiliation{Astrophysics Science Division, NASA Goddard Space Flight Center, 8800 Greenbelt Rd, Greenbelt, MD 20771, USA}

\author[0000-0003-1581-7825]{Ray A. Lucas}
\affiliation{Space Telescope Science Institute, 3700 San Martin Drive, Baltimore, MD 21218, USA}

\author[0000-0001-8688-2443]{Elizabeth J.\ McGrath}
\affiliation{Department of Physics and Astronomy, Colby College, Waterville, ME 04901, USA}

\author[0000-0002-9373-3865]{Xin Wang}
\affil{School of Astronomy and Space Science, University of Chinese Academy of Sciences (UCAS), Beijing 100049, China}
\affil{National Astronomical Observatories, Chinese Academy of Sciences, Beijing 100101, China}

\author[0000-0003-3735-1931]{Stijn Wuyts}
\affiliation{Department of Physics, University of Bath, Claverton Down, Bath BA2 7AY, UK}

\begin{abstract}
We present the data release and data reduction process for the Epoch 1 NIRCam observations for the Cosmic Evolution 
Early Release Science Survey (CEERS).  These data consist of NIRCam imaging in six broadband filters (F115W, F150W, F200W,
F277W, F356W and F444W) and one medium band filter (F410M) over four pointings, obtained in
parallel with primary CEERS MIRI observations (Yang et al. in prep). We reduced the NIRCam imaging with 
the \jwst\ Calibration Pipeline, with custom modifications and reduction 
steps designed to address additional features and challenges with the data.
Here we provide a detailed description of each step in our reduction and a discussion of future expected improvements. Our reduction process includes corrections for known pre-launch issues such as $1/f$ noise, as well as in-flight issues including snowballs, wisps, and astrometric alignment. Many of our custom reduction processes were first developed with pre-launch simulated NIRCam imaging over the full 10 CEERS NIRCam pointings.  We present a description of the creation and reduction of this simulated dataset in the Appendix.
We provide mosaics of the real images in a public 
release, as well as our reduction scripts with detailed explanations 
to allow users to reproduce our final data products.  These represent one of the first official public datasets released from the Directors Discretionary Early Release Science (DD-ERS) program.
\end{abstract}

\section{Introduction} \label{sec:intro}

The Directors Discretionary Early Release Science (DD-ERS) programs are providing the community with early 
and efficient demonstrations of the capabilities of \jwst\ \citep{gardner2006}.
These public programs are designed to test multiple instruments and 
observing modes, and the teams have committed to sharing data products, 
tools and software, simulations and documentation before the \jwst\ Cycle 2 call for proposals. The thirteen ERS programs
cover a wide range of science topics and together have provided a wealth 
of information about early \jwst\ performance and calibration.

Among these programs, the Cosmic Evolution Early Release Science Survey 
(CEERS; ERS 1345, PI: S Finkelstein) is obtaining imaging and spectroscopy of
the Extended Groth Strip (EGS) \textit{Hubble Space Telescope} (\hst) legacy field with three \jwst\ instruments
and five coordinated parallel observing modes.  
The complete program involves imaging with the Near Infrared Camera 
\citep[NIRCam;][]{rieke2003,rieke2005,beichman2012} short and long-wavelength channels in ten pointings, observed as coordinated 
parallels to primary observations with the Near Infrared Spectrograph
\citep[NIRSpec;][]{jakobsen2022} and the Mid-Infrared Instrument
\citep[MIRI;][]{rieke2015,wright2015}. Four of the ten pointings will be additionally covered by NIRCam wide field slitless spectroscopy (WFSS) with 
MIRI imaging obtained in parallel.
The complex CEERS survey layout results in multiple sets of overlapping 
observations, allowing for cross-instrument comparisons and enabling 
explorations of galaxies from $\sim$1 -- 20\micron.
The EGS field is additionally covered by an extensive set of multi-wavelength
observations including imaging and spectroscopy with ground-based observatories, 
\hst, and the \textit{Spitzer Space
Telescope}, from surveys including the Cosmic Assembly Near-infrared Deep Extragalactic 
Legacy Survey \citep[CANDELS;][]{grogin2011,koekemoer2011}, the MOSFIRE Deep Evolution Field Survey
\citep[MOSDEF;][]{kriek2015} and the DEEP2 Galaxy Redshift Survey \citep{newman2013}. 

The CEERS survey is optimized for the study of galaxies in the early 
universe ($z >$ 10) and the processes of galaxy assembly and black hole growth 
for redshifts in the range $z\sim1-10$. The NIRCam and MIRI imaging will 
provide number counts of $z>10$ candidate galaxies, robust stellar 
mass estimates for galaxies at $z>4$, probes of dust-obscured star formation 
and super massive black hole accretion at $z\sim1-3$, and detailed measurements
of source morphologies. 
Multi-object spectroscopy with NIRSpec's micro-shutter assembly (MSA) and 
NIRCam's wide field slitless spectroscopy (WFSS) will provide spectroscopic
redshifts of sources at $z \sim 5-12$ through rest-ultraviolet (UV) and/or rest-optical 
emission lines, helping to constrain models of chemical evolution in the 
interstellar medium.
The full details of the CEERS science goals and observing strategy will
be presented in Finkelstein et al. (in prep).  

In this paper, we present the CEERS team reduction of our NIRCam imaging,
providing a detailed description of our processing steps with the 
\jwst\ Calibration Pipeline. Building on our team's extensive experience 
from \textit{HST}'s CANDELS program we have developed a number of custom processing steps and modifications to the Calibration Pipeline that improve on aspects such as 
imaging noise, detector features, alignment and global background modeling.
Alongside this paper we will announce the first public release of our NIRCam images, available for download at \url{ceers.github.io/releases.html} and on MAST as High Level Science Products via \dataset[10.17909/z7p0-8481]{\doi{10.17909/z7p0-8481}}. We include in this 
release scripts for downloading the raw NIRCam imaging and processing it 
through the Calibration Pipeline along with all our custom scripts.

As part of our preparations for reducing the real NIRCam imaging, we created extensive simulations to replicate the CEERS observing strategy and expected
detector noise and artifacts as closely as possible. These simulations help
validate methods for source detection, photometry and morphological 
measurements, and were instrumental in building our strategies for working 
with the real NIRCam data. We have previously released our simulations 
as part of the CEERS Simulated Data Releases, which are available at the 
same website, and have also included NIRSpec MSA, NIRCam WFSS, and MIRI 
imaging simulations. 

\begin{figure*}
\epsscale{0.9}
\plotone{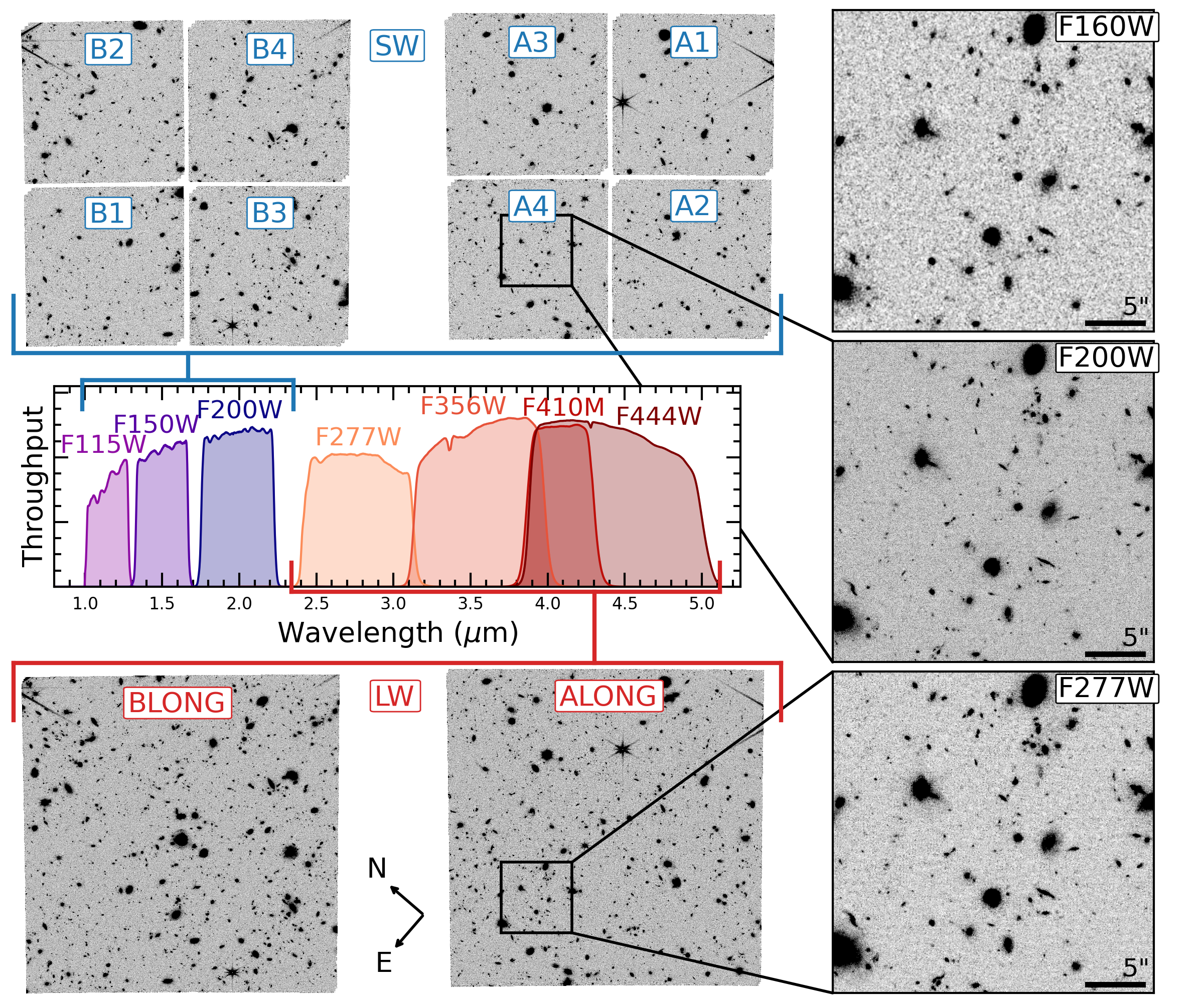}
\caption{An illustration of the CEERS NIRCam observations, using NIRCam1 as an example. We show a SW (F200W) mosaic on the top left and a LW (F277W) mosaic on the bottom left. The SW and LW detectors are labeled in blue and red, respectively. The large module gap is visible in both mosaics, and the SW detector gaps are visible in the F200W mosaic. The CEERS filter throughput curves are plotted in the middle, with SW filters shown in purple-blue and LW filters in orange-red. Along the right edge, we show $29\times29$\arcsec\ zoom-ins of the mosaic, with the same portion of the \hst\ F160W mosaic displayed on top for reference. The black bars in the lower right corners of the zoom-in panels indicate a 5\arcsec scale.
\label{fig:nircam}}
\end{figure*}

This paper is organized as follows.
In Section~\ref{sec:obs}, we describe the CEERS NIRCam observations obtained
to date. We describe our image reduction in detail in 
Section~\ref{sec:realreduction}, presented in three stages (Sections~\ref{sec:stage1}, \ref{sec:stage2}, and \ref{sec:stage3}). We discuss our custom corrections to 
remove snowballs (Section~\ref{sec:snowballs}), wisps (Section~\ref{sec:wisps})
and $1/f$ noise (Section~\ref{sec:1fnoise}). In Section~\ref{sec:stage3}, 
we present the processing steps required to create combined mosaics, including 
astrometric alignment (Section~\ref{sec:astrometry}), resampling
(Section~\ref{sec:mosaics}) and a custom background subtraction
(Section~\ref{sec:background}). We discuss known issues present in our 
reduced data products in Section~\ref{sec:issues}, and briefly summarize in Section~\ref{sec:summary}. We also provide a 
detailed appendix, with a description of our released mosaic images
(Appendix~\ref{sec:files}), brief summaries of previous NIRCam data reduction 
versions that were used in early publications by the CEERS team 
(Appendix~\ref{sec:prevreduc}), and a comprehensive discussion of the
creation and reduction of our simulated NIRCam imaging 
(Appendices~\ref{sec:sims} and \ref{sec:simsreduction}).
We express all magnitudes in the AB system \citep{oke1983}
unless otherwise noted.

\section{CEERS Epoch 1 Observations} \label{sec:obs}
Due to the observabilty of the EGS field and the need for self-overlap of the CEERS observing modes, CEERS is schedulable in June or December, with a 180$^{\circ}$
position angle rotation between these times achieving similar survey layouts. Due to time constraints in the June 2022 observing 
window following telescope and instrument commissioning (EGS observability ends on July 1), the CEERS 
observations were split into two epochs. 
The first epoch was executed on 21 June, 
2022. This involved CEERS pointings 1, 2, 3, and 6, which were observed with 
MIRI as the primary instrument and NIRCam in parallel. The remaining
CEERS observations (six NIRSpec MSA+NIRCam imaging pointings and four NIRCam
WFSS+MIRI imaging pointings) are scheduled for December 2022. We note that the current MIRI and NIRCam observations do not overlap, but many of the MIRI pointings will have NIRCam coverage once the rest of the survey observations are completed in December.
In this paper, we focus on the CEERS Epoch 1 NIRCam observations  
(hereafter referred to as NIRCam1, NIRCam2, NIRCam3, and NIRCam6) and 
reduction. 

\begin{deluxetable}{cccc}
\tablecaption{CEERS Epoch 1 NIRCam Observations\label{tab:obs}}
\tabletypesize{\small}
\tablehead{
\multicolumn2c{Filters} & \colhead{$N$ Groups} & \colhead{Exptime} \\
\colhead{NIRCam} & \colhead{MIRI} & \colhead{} & \colhead{(s)}}
\startdata
\multicolumn{4}{c}{NIRCam1 Field Center: 14:19:56.2 +52:58:38.8} \\
F115W+F277W & F770W & 5 & 1546.1 \\
F115W+F277W & F1000W & 5 & 1546.1 \\
F115W+F356W & F1280W & 5 & 1546.1 \\
F115W+F356W & F1500W & 5 & 1546.1 \\
F150W+F410M & F1800W & 5 & 1546.1 \\
F150W+F410M & F2100W & 5 & 1546.1 \\
F200W+F444W & F2100W & 9 & 2834.5 \\
\hline
\multicolumn{4}{c}{NIRCam2 Field Center: 14:19:34.8 +52:54:50.3} \\
F115W+F277W & F770W & 5 & 1546.1 \\
F115W+F277W & F1000W & 5 & 1546.1 \\
F115W+F356W & F1280W & 5 & 1546.1 \\
F115W+F356W & F1500W & 5 & 1546.1 \\
F150W+F410M & F1800W & 5 & 1546.1 \\
F150W+F410M & F2100W & 5 & 1546.1 \\
F200W+F444W & F2100W & 9 & 2834.5 \\
F200W+F444W\tablenotemark{a} & F2100W & 9 & 2834.5 \\
\hline\multicolumn{4}{c}{NIRCam3 Field Center: 14:19:12.7 +52:51:03.5} \\
F115W+F277W & F560W & 9 & 2834.5 \\
F115W+F356W & F770W & 9 & 2834.5 \\
F150W+F410M & F770W & 9 & 2834.5 \\
F200W+F444W & F770W & 9 & 2834.5 \\
\hline\multicolumn{4}{c}{NIRCam6 Field Center: 14:19:25.2 +52:49:56.0} \\
F115W+F277W & F560W & 9 & 2834.5 \\
F115W+F356W & F770W & 9 & 2834.5 \\
F150W+F410M & F770W & 9 & 2834.5 \\
F200W+F444W & F770W & 9 & 2834.5 \\
\enddata
\tablenotetext{a}{This observation was repeated at a slightly different 
PA on 28 June 2022 due to a problem with the MIRI observations, but is heavily affected by persistence, see Section~\ref{sec:ceers2b} for details.}\label{note:obstab}
\tablecomments{All images are obtained in parallel with MIRI imaging as
the primary instrument mode. All exposures use the MEDIUM8 readout
pattern and a 3-point dither pattern determined by the MIRI PSF. The exposure times listed are the total including dithers.
}
\end{deluxetable}

The NIRCam instrument \citep{rieke2003,rieke2005,beichman2012} consists of two modules, A and B, each of which 
have a field-of-view covering $2.2\arcmin \times 2.2\arcmin$. The modules
are separated by a $\sim$45\arcsec\ gap, making the total combined NIRCam
field-of-view $\sim$9.7 arcmin$^2$. Each module has four short wavelength 
(SW) detectors, A1-A4 and B1-B4, tuned for observations in the range
$0.6-2.3$\micron\ and one long wavelength (LW) detector (ALONG and BLONG,
$2.4-5$\micron). 
Together, the SW detectors cover approximately the same area as the LW detectors, though they are separated by $\sim$5\arcsec, leaving gaps in the SW mosaics that are not present in the LW mosaics.
Observations in the
SW and LW channels are obtained simultaneously. For CEERS Epoch 1, we have 
paired the following SW and LW filters together: F115W+F277W,
F115W+F356W, F150W+F410M, and F200W+F444W. The filters are observed in this 
order to ensure that fading persistence from previous observations will
not mimic a Lyman break in our images.  
In Figure~\ref{fig:nircam}, we show the seven NIRCam filters as well as
a SW (F200W) and LW (F277W) mosaic of NIRCam1, with each individual detector
labeled. The primary MIRI imaging is obtained with seven filters (see Table~\ref{tab:obs}): F560W, F770W, F1000W, F1280W, F1500W, F1800W and F2100W. The CEERS MIRI imaging and reduction will be presented in Yang et al. (in prep).

We use a three-point dither pattern chosen to optimize the subpixel sampling 
and bad pixel mitigation for both NIRCam and primary MIRI observations. As
MIRI dithers are required to be $>3\times$ the full width at half maximum
(FWHM) of the MIRI point spread function (PSF), the dither steps are 
determined by the MIRI filter for each observation. Specifically, we 
use the dithers \texttt{3-POINT-MIRI-[filter]-WITH-NIRCam}, where 
\texttt{filter} is the primary MIRI filter. (The exception is for the observations with MIRI filter F560W, for which a custom dither pattern was not available and so we adopt that for F770W.) As a result, different dither step 
sizes are sometimes used for images in the same NIRCam filter. However,
as illustrated in Figure~\ref{fig:nircam}, most of the dithers are not
large enough to cover the SW detector gaps. Our NIRCam exposures use the 
MEDIUM8 readout pattern with either five or nine groups and one integration. 
The total exposure time is $\sim$3000 seconds in each filter, with double 
the exposure time in F115W for added depth in the \lya\ dropout filter 
for $z\gtrsim9.5$ galaxies. Exposure times for each observation are listed in Table~\ref{tab:obs}.

\begin{deluxetable*}{llc}
\tablecaption{CEERS NIRCam v0.5 Reduction\label{tab:reduction}}
\tabletypesize{\small}
\tablehead{
\colhead{Reduction Step} & \colhead{Script Name} & \colhead{Discussion in Text}}
\startdata
Stage 1 + Snowball Correction & \texttt{snowball\_wrapper.py} & Sections~\ref{sec:stage1}, \ref{sec:snowballs} \\
Wisp Subtraction & \texttt{wispsub.py} & Section~\ref{sec:wisps}\\
$1/f$ Noise Removal & \texttt{remstriping.py} & Section~\ref{sec:1fnoise} \\
Stage 2 & \texttt{image2\_1.7.2.asdf} & Section~\ref{sec:stage2} \\
SkyMatch + Outlier Detection & \texttt{image3\_part1.asdf} & Section~\ref{sec:stage3} \\
Astrometric Alignment & \texttt{run\_tweakreg.py} & Section~\ref{sec:astrometry} \\
Individual Background Subtraction + Variance Map Rescaling & \texttt{skywcsvar.py} & Section~\ref{sec:mosaics} \\
Mosaic Creation & \texttt{image3\_nircam[pointing]\_[chn].asdf\tablenotemark{a}} & Section~\ref{sec:mosaics} \\
Mosaic background subtraction & \texttt{mosaic\_background.py} & Section~\ref{sec:background} \\
\enddata
\tablenotetext{a}{We use different parameter files for creating SW and LW mosaics, where the only difference is the input-to-output pixel size ratio. Here, \texttt{chn} refers to either the SW or LW channels.}
\end{deluxetable*}

Finally, due to an on-board problem with MIRI pointing 2 at the time of 
observation, the last set of MIRI+NIRCam exposures in this pointing was
re-observed on 28 June 2022 (see footnote~\textit{a} in Table~\ref{tab:obs}). The original NIRCam observations were not affected.
As a result, there is a second set of F200W+F444W imaging in this field at 
a slightly different position angle, providing added depth in these two 
filters. We note, however, that these additional images seem to be heavily 
affected by persistence (see Section~\ref{sec:ceers2b} for a discussion). 
See Table~\ref{tab:obs} for details on Epoch 1 NIRCam observations
and Finkelstein et al. (in prep) for more information on the CEERS Survey and 
observation specifications. 

\section{Image Reduction} \label{sec:realreduction}
We process the raw images through the \jwst\ Calibration Pipeline 
created and maintained by STScI (hereafter \pipeline), 
with custom steps and modifications informed from our work with the 
simulations described in Appendix~\ref{sec:simsreduction}. The reduction described here and used to create our first public data release uses 
\pipeline\ version 1.7.2 and Calibration Reference Data System pipeline mapping (CRDS\footnote{\url{jwst-crds.stsci.edu} \label{note:crds}} pmap) 0989, which 
includes in-flight NIRCam dark, distortion, bad pixel mask, readnoise, and 
superbias reference files.
It also includes the ground flats corrected for in-flight performance 
released on 18 August 2022 (pmap 0951) and the updated photometric calibration 
reference files\footnote{\url{www.stsci.edu/contents/news/jwst/2022/an-improved-nircam-flux-calibration-is-now-available}} released on 4 October 2022.
Pipeline mapping 0989 corresponds to the NIRCam instrument mapping (imap) 
0232. 

In the following sections we describe our image reduction steps in detail. 
We will also provide the parameter files for each pipeline step, custom Python 
routines for the additional corrections we have applied, and batch scripts 
for running all steps on all four NIRCam pointings on GitHub shortly following
the data release.
These scripts are summarized in Table~\ref{tab:reduction} for reference. 
We invite the reader to use these scripts and parameter files to reproduce 
our reduction.

\begin{figure}
\epsscale{1.1}
\plotone{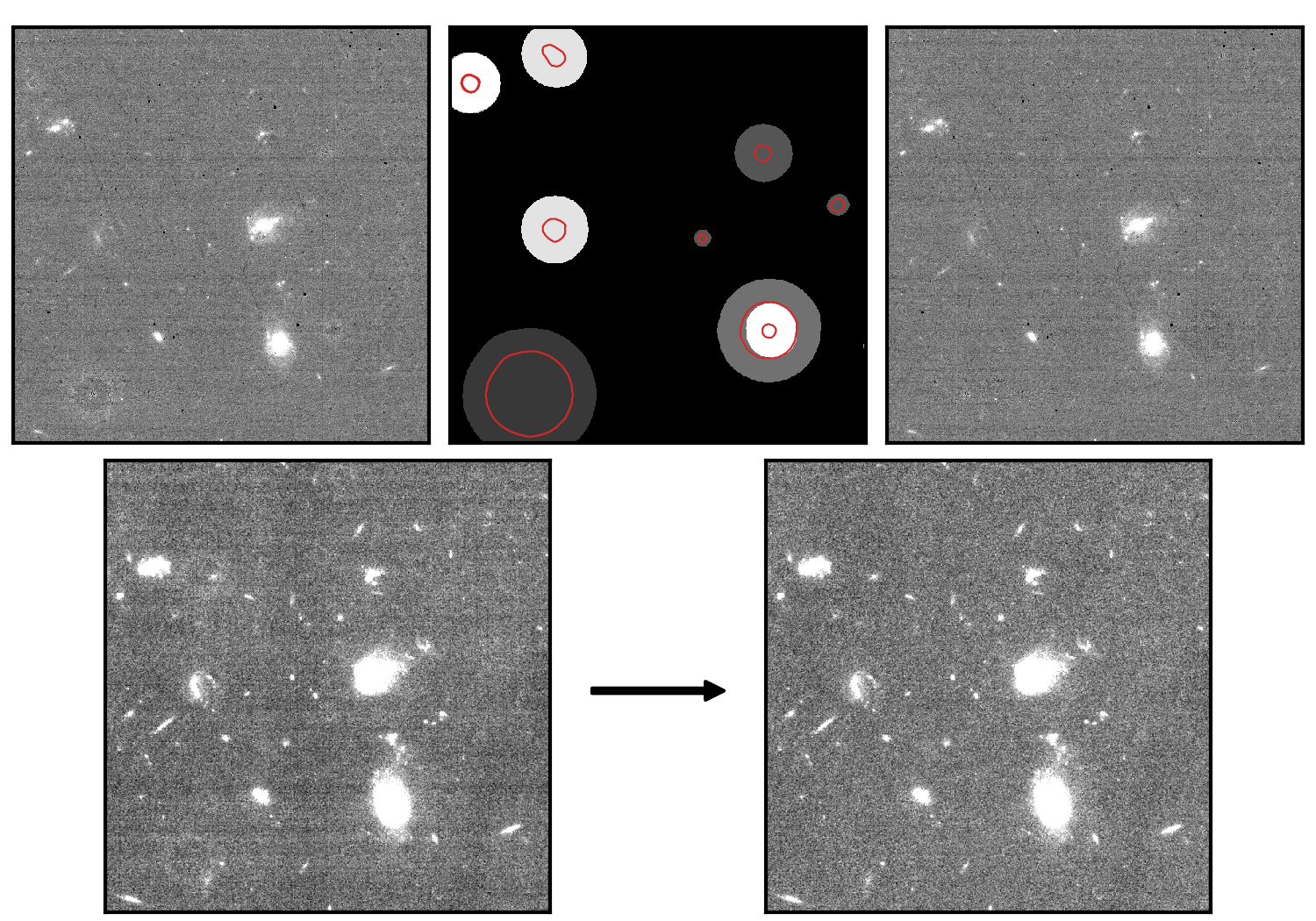}
\caption{An example of the snowball correction. In the top row we show 
a portion of an F115W countrate map that includes several snowballs. 
The left image shows the output of the default \pipeline, where there is 
a particularly large snowball present in the lower left corner of the image.
The middle shows the mask generated by identifying and growing the snowball
footprints, where the red contours identify the original footprints. The
snowballs are color-coded by the group in which they were detected, brighter
snowballs were detected in later groups. The right panel shows the countrate 
map resulting from refitting the ramp with the updated snowball mask added 
to the \texttt{GROUPDQ} arrays. In the bottom row we show the portion of 
the F115W mosaic that corresponds to this region. In the left panel, the 
mosaic has been created without correcting for the snowballs in the 
individual exposures, while the right panel shows a mosaic that included 
this correction. The three CEERS dithers are not sufficient to compensate 
for the presence of snowballs. This additional correction is needed to remove them from the coadded mosaics. We 
note that the left panel also includes $1/f$ noise, which we discuss in
Section~\ref{sec:1fnoise}.
\label{fig:snowballs}}
\end{figure}

\subsection{Stage 1 -- Detector-level corrections}\label{sec:stage1}
Stage 1 of the \jwst\ Calibration Pipeline performs detector-level corrections,
many of which are common to all instruments and observing modes. The reduction 
steps involve initializing the data quality (DQ) arrays for flagging pixels;
identifying saturated pixels; subtracting the superbias; 
using reference pixels to correct for readout noise;
correcting pixels for non-linearity;
subtracting the dark current; 
identifying cosmic rays as jumps in each pixel's up-the-ramp signal;
and calculating a linear fit to the unflagged ramp data to determine the 
average countrate per pixel. 
The end product of Stage 1 is a countrate image in units of counts/s.
We adopt the default parameter values for these steps. We run the reduction 
steps of Stage 1 together with the snowball correction described in the 
following section (\ref{sec:snowballs}), using the Python script 
\texttt{snowball\_wrapper.py} described in the following section.
For reference we also provide the pipeline parameter file 
\texttt{detector1\_1.7.2.asdf} with the equivalent set of default parameter 
values.

In the following subsections, we describe some features of the NIRCam images
for which we have developed custom corrections. We refer the reader to 
\citet{rigby2022} for more information on detector performance and some of 
these features that were discovered and characterized during commissioning.

\subsubsection{Snowball Correction} \label{sec:snowballs}
``Snowballs'' are large cosmic ray events that can affect hundreds of pixels
and have a circular morphology on the NIRCam detectors. 
We see an average of $\sim25-30$ snowballs per detector in a 
$\sim$900 second exposure. 
The fluence of counts in the center of the snowballs is
very high and can sometimes saturate the detector. The cosmic ray flagging 
step of \pipeline\ often successfully identifies and flags the central 
cores of these large cosmic rays, leaving the more diffuse wings of the 
snowballs present in the count rate maps output by Stage 1.
The top left panel of Figure~\ref{fig:snowballs} shows an example of 
a countrate map containing multiple snowballs that were only partially 
removed by the pipeline.

We identify snowballs as large contiguous sets of pixels in the 
\texttt{GROUPDQ} arrays that have been flagged as jumps (\texttt{JUMP\_DET}). We found it helpful to divide the snowballs into two tiers: large ones, which require a rather large mask to address the extended wings, and smaller ones, which do not require masking as much additional area. In order to separate snowballs from smaller cosmic ray impacts, we median 
filter the \texttt{GROUPDQ} arrays with separate two-dimensional tophat kernels 
with radii of 7 and 15 pixels to identify the smaller and larger snowballs, respectively. This filtering recovers extended groups of 
pixels that have an approximately circular footprint. We also include in the 
snowball mask any saturated pixels that are within these big groups. We then 
grow the resulting snowball footprints via binary dilation and a two-stage 
tophat growing kernel with radii of 7 and 35 pixels. We add this updated snowball
mask to the \texttt{GROUPDQ} array, and run the ramp-fitting step from 
\pipeline\ Stage 1. The flux in the affected pixels is determined by the 
remainder of the ramp, using the slope of the ramp excluding the newly-identified
cosmic ray jump. 
The top middle panel of Figure~\ref{fig:snowballs} shows the DQ array for 
the count rate map in the left panel, with the original footprints of the 
snowballs identified with red contours. The top right panel shows the output
countrate map that results from performing this snowball correction.
In the bottom row, we show the portion of the mosaic that corresponds to 
the region displayed in the top row to demonstrate the effectiveness of 
this correction.

\begin{figure*}
\plotone{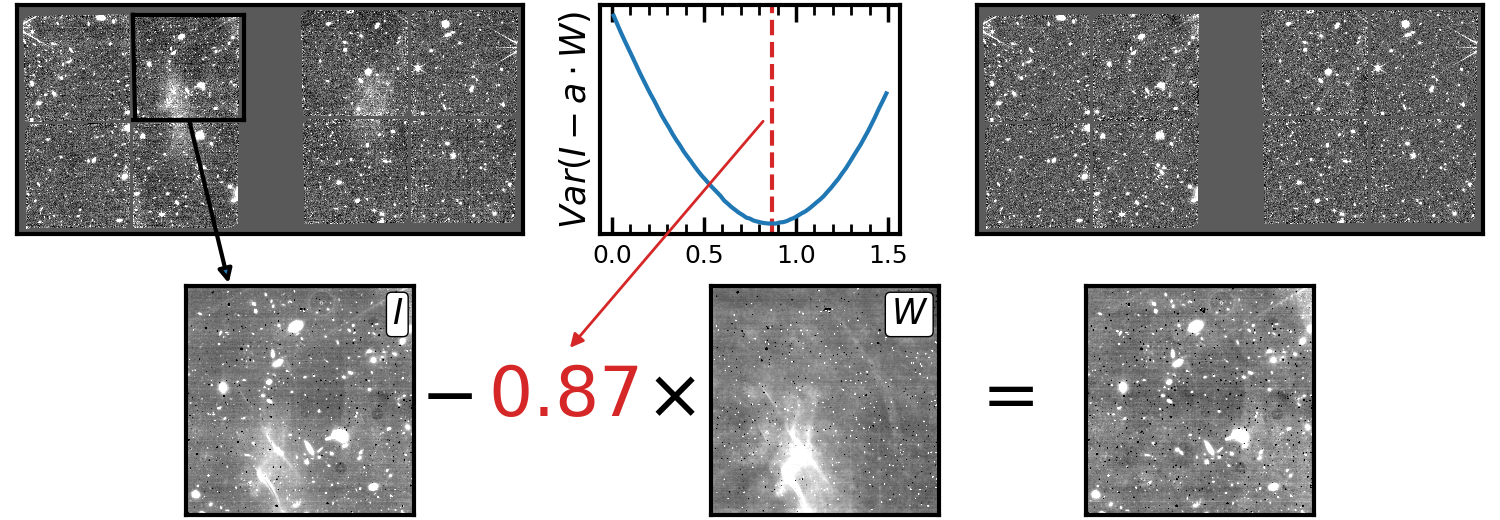}
\caption{
An illustration of our wisp subtraction process. The upper left panel shows the 
full extent of the wisps in an F200W mosaic of CEERS NIRCam pointing 1. We 
then illustrate the process of fitting the wisp templates using one F200W 
image from detector B4 as an example. We minimize the variance of the wisp 
template $W$ scaled by a factor $a$ and subtracted from the image $I$ in 
order to find the best-fitting $a$ (0.87 in this case). The lower right panel
shows a cleaned version of the F200W image where the wisp feature has been 
removed. The upper right panel shows the full F200W mosaic for this pointing
once the wisps are subtracted from all affected input images.  
The three images in the bottom row have been smoothed by a Gaussian 
kernel with $\sigma=2$ pixels for display purposes.
\label{fig:wisps}}
\end{figure*}

The snowball correction is performed with our Python script 
\texttt{snowball\_wrapper.py}, that runs Stage 1 of the pipeline, saves the 
ramps from the first run of the ramp fitting step, identifies the snowballs
and grows their footprint, flags them as cosmic rays, and then runs the 
ramp fitting step again to create a countrate map that excludes the 
flagged portions of the ramps.

We note that persistence from the saturated cores of very bright snowballs can show up in subsequent exposures. These are faint enough not to be rejected when the dithered images are combined. They can show up as single-band detections that look a lot like faint emission-line galaxies. For pure emission-line sources in WFSS exposures, it is important to inspect the individual exposures and their data-quality arrays.

\subsubsection{Wisp Subtraction} \label{sec:wisps}
We next subtract the ``wisp'' features from the F150W and F200W images. 
Wisps are created from stray light reflected off the secondary mirror
supports. They are visible on detectors A3, A4, B3 and B4, and are most 
prominent in F150W and F200W. The strength of the wisp features depends on 
the source of the reflected light, and so the wisps can have a variable 
brightness from exposure to exposure. The top left panel of 
Figure~\ref{fig:wisps} shows a version of the F200W mosaic for CEERS NIRCam 
pointing 1 from which we have not subtracted the wisps as an example of the 
large scale structure of this feature. They extend across two detectors in 
each module.

The NIRCam team has provided wisp templates constructed from images obtained 
by several NIRCam commissioning and ERS programs observed early in Cycle 1.
At the time of this writing, two sets of templates have been released, in July and August 2022. Both sets of templates are available on the \jwst\ User Documentation page 
on Claws and Wisps\footnote{The updated templates are packaged in wisps\_2022\_08\_26.tgz available at \url{jwst-docs.stsci.edu/jwst-near-infrared-camera/nircam-features-and-caveats/nircam-claws-and-wisps}}. We use the updated wisp 
templates released 26 August 2022, which are the result of reprocessing the original 
observations with updated processing steps and reference files, including
a correction for the large-scale variations in the ground flats consistent 
with that introduced by pmap 0956.

We scale the templates to account for the variable brightness of the wisp 
feature and subtract them from the images in the following way. 
We first apply the flat field to each image to match the flat-fielded templates
and to measure the wisp feature in the flattened images.
For each image, we perform a very ``cold'' source detection (i.e., a very high detection threshold of 5.5$\sigma$) with \photutils\ \citep{bradley2020} 
 to identify and mask 
out large, bright sources in the image while avoiding the wisp feature itself. 
Next, we smooth the wisp template with a Gaussian kernel with $\sigma=2$~pixels.
For an array of coefficients $a$, we find the value that minimizes 
$\sigma_{MAD}^2(I - a W)$, where $\sigma_{MAD}$ is the  median absolute
deviation, $I$ is the flat-fielded masked image, and $W$ is the smoothed wisp 
template. We find that in almost all cases, we need to scale the wisp 
template by a factor $<1$.
We scale the original (unsmoothed) wisp template by $a$ and subtract it from
the original (unflatfielded and unmasked) image. 
We illustrate this process in Figure~\ref{fig:wisps}.

\begin{figure*}
\plotone{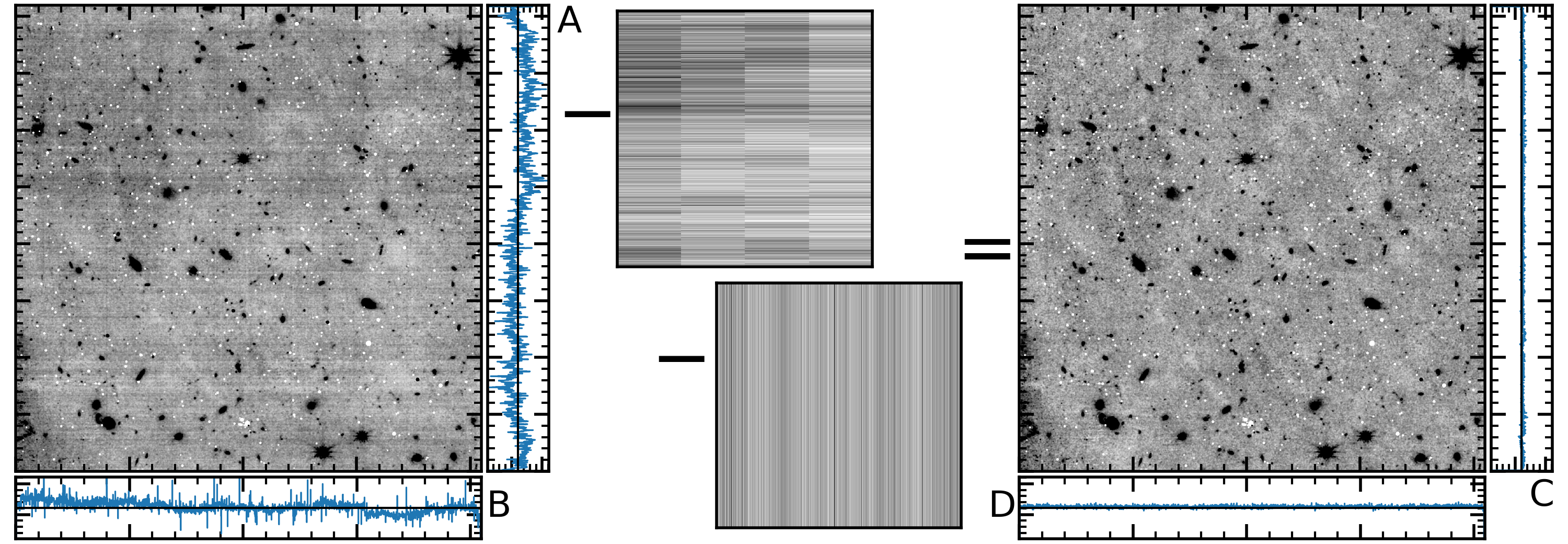}
\caption{
An illustration of our process for removing $1/f$ noise from countrate maps.
The image on the left shows an input countrate map in F200W affected by $1/f$ noise. 
We show the median pixel value measured in the unmasked portions of each row
(panel A) and column (B) in blue along the right and bottom edges, 
respectively, showing significant variation due to 1/f noise. The middle panels show the horizontal and vertical striping
patterns that we identify using the unmasked portions of the image. The
amplifier-dependent pattern is visible in the horizontal striping model. The
image on the right shows the result of subtracting both models, where the 
right (C) and bottom (D) panels show that the noise in the image has been
significantly reduced. All four panels displaying the median pixel values 
(A-D) are plotted with the same limits $\pm0$.
The countrate maps are flat-fielded and smoothed by a Gaussian kernel 
with $\sigma=2$ pixels.
\label{fig:1fnoise}}
\end{figure*}

While this scaling does well at removing the large-scale feature, our wisp 
subtraction is still preliminary. The current wisp templates were produced 
with an updated but still early NIRCam reduction. Additionally, 
in early Cycle 1 observations, the dithers were often small compared to the 
size of the sources observed. As a result, some ghost images of 
bright sources are present in some of the wisp templates and therefore 
become regions of oversubtraction (by $\sim3-4$\%) in the CEERS images.
We also note that we see some evidence for the presence of weak 
wisps in F115W, but as there are no templates available for this filter, 
we leave it uncorrected.
The wisp templates will be improved throughout Cycle 1 with additional 
observations and updates to the NIRCam reduction and reference files.

\subsubsection{1/f Noise Subtraction} \label{sec:1fnoise}

Our final custom step on the countrate images is to measure and remove 
\textit{1/f} noise, which is correlated noise introduced in the images when 
the detectors are read out \citep{schlawin2020}. The noise presents as 
horizontal and vertical striping patterns that vary from row to row and 
column to column. Ideally, this noise would be fit and removed during the 
reference pixel fitting or the up-the-ramp fitting of \pipeline, an update that 
may be included in future versions of the Calibration Pipeline. In the 
meantime, we perform our own correction for \textit{1/f} noise in each
individual countrate map as follows. 

As an additive effect, the noise pattern should be removed before applying 
the multiplicative flat field correction, yet the striping patterns are best 
measured on flat images. We therefore apply the flat field to the countrate
maps for pattern measurement, but we subtract it from the original countrate
images. We next mask all bad pixels (with a data quality flag value $>$0)
and source flux that is identified with \photutils\
using a tiered source detection method. This method consists of convolving 
the image with progressively smaller kernels and running a segmentation-style 
source detection at each step. In this way we are able to detect both large,
extended sources and small, compact sources with detection parameters 
optimized for each source size. 
We use four tiers, with Gaussian kernels of $\sigma=25$, 15,
5 and 2 pixels (on the original 0\farcs031/pixel and 0\farcs063/pixel scales
for the SW and LW channels, respectively). These values were chosen after
experimenting with several filter kernels to aggressively mask as much source
flux as possible. Next we calculate a pedestal value for the sky background
by fitting a Gaussian to the distribution of pixel values in the masked
image. 

We measure the striping pattern using a sigma-clipped (2$\sigma$) median along first 
rows and then columns. For the row correction (horizontal striping), we 
measure and remove the pattern amplifier-by-amplifier. In some images,
especially in the SW filters, the difference from amplifier to amplifier 
can vary by $\sim3-5$\%, and so this amplifier-dependent correction works better than using the median from the full row. However, in some cases, especially around bright or extended sources, a large 
number of pixels in a given amp-row are masked leaving too few to 
calculate a robust median. In these cases, we use the median of the 
entire row for the affected amp-row. The threshold used to determine the
cutoff in number of masked pixels per amp-row varies from image to image to allow for the best pattern subtraction.
We illustrate this amp-row and column subtraction in Figure~\ref{fig:1fnoise}.

This amp-by-amp approach works well for NIRCam imaging, especially 
for fields like the EGS that are relatively sparsely populated (compared to a globular 
cluster, star field, or regions with high nebulosity, for example, where a majority of pixels in individual amplifier rows will be masked). However,
it is not a good approach for NIRCam WFSS observations, 
where the length of a trace in the row grism is approximately the same size
as an amplifier for some filters.

\subsubsection{Additional Imaging in NIRCam2}\label{sec:ceers2b}
\begin{figure}
\plotone{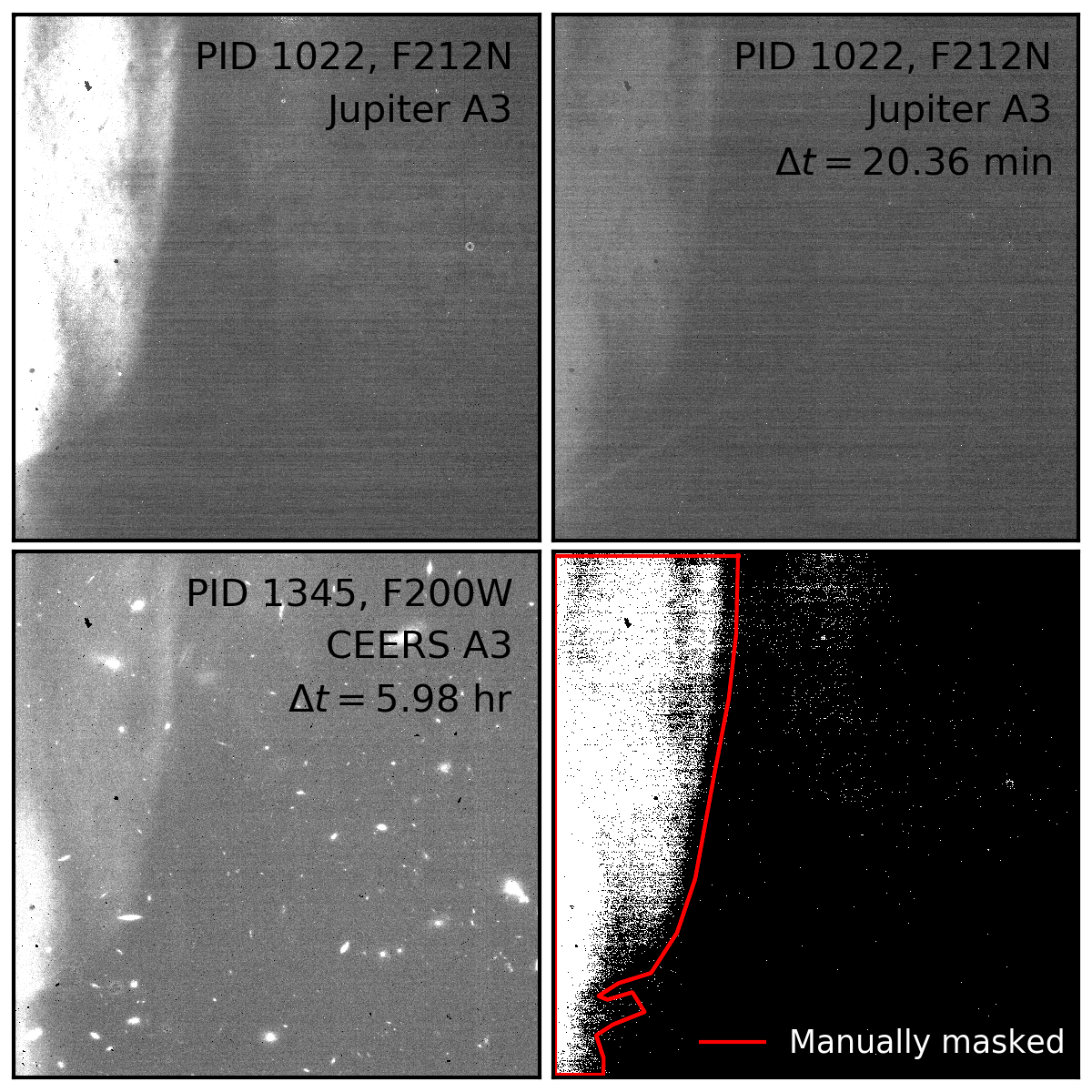}
\caption{The persistence in the additional F200W imaging in NIRCam2. The top row shows persistent or scattered light present in A3 images of commissioning program 1022, in which Jupiter was observed on detector A1. These two images are displayed with the same zscale stretch to highlight how the illumination pattern is fading with time. One of the CEERS F200W exposures obtained $\sim$6 hours later is displayed in the lower left panel, where the same pattern is visible. We constructed a mask of pixels with elevated counts in both images from program 1022, shown in the lower right panel. We then manually masked the pixels within the red histogram, necessary to mask the fainter edges of the pattern that otherwise are present in the final mosaic. None of the other detectors contain clear signatures of persistence. Regardless, a future reduction will include a more careful treatment of persistence in these observations.
\label{fig:ceers2b}}
\end{figure}

The observations of CEERS MIRI pointing 2 on 21 June 2022 encountered a
problem writing data to disk, and one of the F2100W (with NIRCam F200W+F444W
in parallel) image sets was repeated on 28 June. See Yang et al. (in prep) for a
full discussion of the effect on the MIRI images.
The NIRCam images from 21 June were not affected, and the repeated observations
offer additional depth in F200W and F444W, though at a slightly different
position angle. However, the repeat observations obtained on 28 June appear
to suffer from significant persistence on SW detector A3.

We searched MAST for all NIRCam observations obtained in the 24 hours before
the NIRCam2 repeat imaging. Program 1022 (PI: J Stansberry) observed Jupiter
$\sim$6 hours before the CEERS imaging. This commissioning program aimed to
test the Fine Guidance Sensor (FGS) guiding near a bright planet as well as
to characterize the scattered light in each instrument.
There were two observations obtained with Jupiter on SW detector A1 (F212N)
and LW detector ALONG (F322W2). In both cases, the same illumination pattern
is present on SW detector A3 as that we see in the CEERS images.
We show the calibrated (processed through pipeline Stages 1 and 2,
flat-fielded and in units of MJy/sr) A3 detector images from program 1022
in the top two panels of Figure~\ref{fig:ceers2b}. The pattern is fainter in
the second image, obtained 20.3 minutes after the first, indicating that the
pattern is fading with time as expected for persistence.
The feature on the detector A3 images may be scattered light from Jupiter      
imaged off-detector, or fading persistence from an earlier observation.
(The program observed before 1022, and 8--10 hours before CEERS, is 
proprietary -- GO 2473, PI: L Albert.)
We note that neither the observations of Jupiter on detector A1 nor any of 
the LW images left persistence in the CEERS images.      

We create a mask using the two A3 images from program 1022, identifying 
affected pixels as those with a flux $>$20 MJy/sr in the first exposure and 
$>$10 MJy/sr in the second exposure. These thresholds were found through trial 
and error to identify the persistent pattern and avoid flagging noise 
fluctuations. The lower right panel of Figure~\ref{fig:ceers2b} shows the mask 
we created in this way, with flagged pixels in white. 
However, we found that this mask was not sufficient around the scattered 
light pattern. The wings of the pattern were still present in our F200W mosaics,
and so we manually masked out a region around the affected area, shown as 
the red polygon in Figure~\ref{fig:ceers2b}. This method represents a
preliminary approach to handling the persistence in this field. In future
data reductions, we will create a more careful mask incorporating the CEERS
images to identify the fading pattern. Here we have also created a mask for
detector A3 only. In the future we will explore persistence on all detectors.

We apply the mask to the snowball-corrected countrate maps by setting the 
image pixels to zero in the science extensions
and setting their data quality (DQ) values to \texttt{D0\_NOT\_USE}.
We then perform the corrections for wisps and $1/f$ noise as described above.
We create three sets of mosaics in F200W and F444W for NIRCam2: one containing
just the original imaging, one with just the repeated imaging, and one
including both sets of imaging.

\subsection{Stage 2 -- Individual Image Calibrations}\label{sec:stage2}
Stage 2 of \pipeline\ involves steps such as flat-fielding the data
and applying the flux calibration that converts the images from counts/s to 
MJy/sr. We adopt the default values for these pipeline steps. 
We note that the NIRCam flats used are ground flats that have been corrected
for an illumination gradient using in-flight observations.
Additionally, the flux calibration is based on observations of three standard 
stars (PIDs 1536, 1537 and 1538) that have been observed across all ten NIRCam 
detectors. It builds on the work of several teams, including the Resolved
Stellar Populations ERS program (PID 1334, PI D. Weisz) observations of the
globular cluster M92 \citep{boyer2022a,nardiello2022}, zeropoints derived by G. 
Brammer\footnote{\url{github.com/gbrammer/grizli/pull/107}}, and early 
absolute flux zeropoints from the NIRCam team (M. Rieke, private communication).
Both the flat fields and the flux calibration tables are important reference 
files that will continue to be updated throughout Cycle 1. 

\subsection{Stage 3 -- Ensemble Processing}\label{sec:stage3}
Stage 3 performs ensemble reduction steps, where the output product is a single
mosaic per filter combining all images and dithers. 
These steps include astrometric alignment, background matching, 
outlier detection, and resampling the images onto a common output grid.
We break this stage up into individual steps, first applying a customized
astrometric correction, then running the \texttt{OutlierDetection} step of
\pipeline\ Stage 3 with the default parameter values.
Then we background-subtract each individual file and perform two
additional corrections on the variance arrays before resampling the 
images into one combined mosaic per filter. Finally, we perform a custom
background subtraction of the mosaics to remove any residual background.
In the following subsections, we describe these customized processing 
steps.

\subsubsection{Astrometric Alignment} \label{sec:astrometry}
\begin{figure*}
\plotone{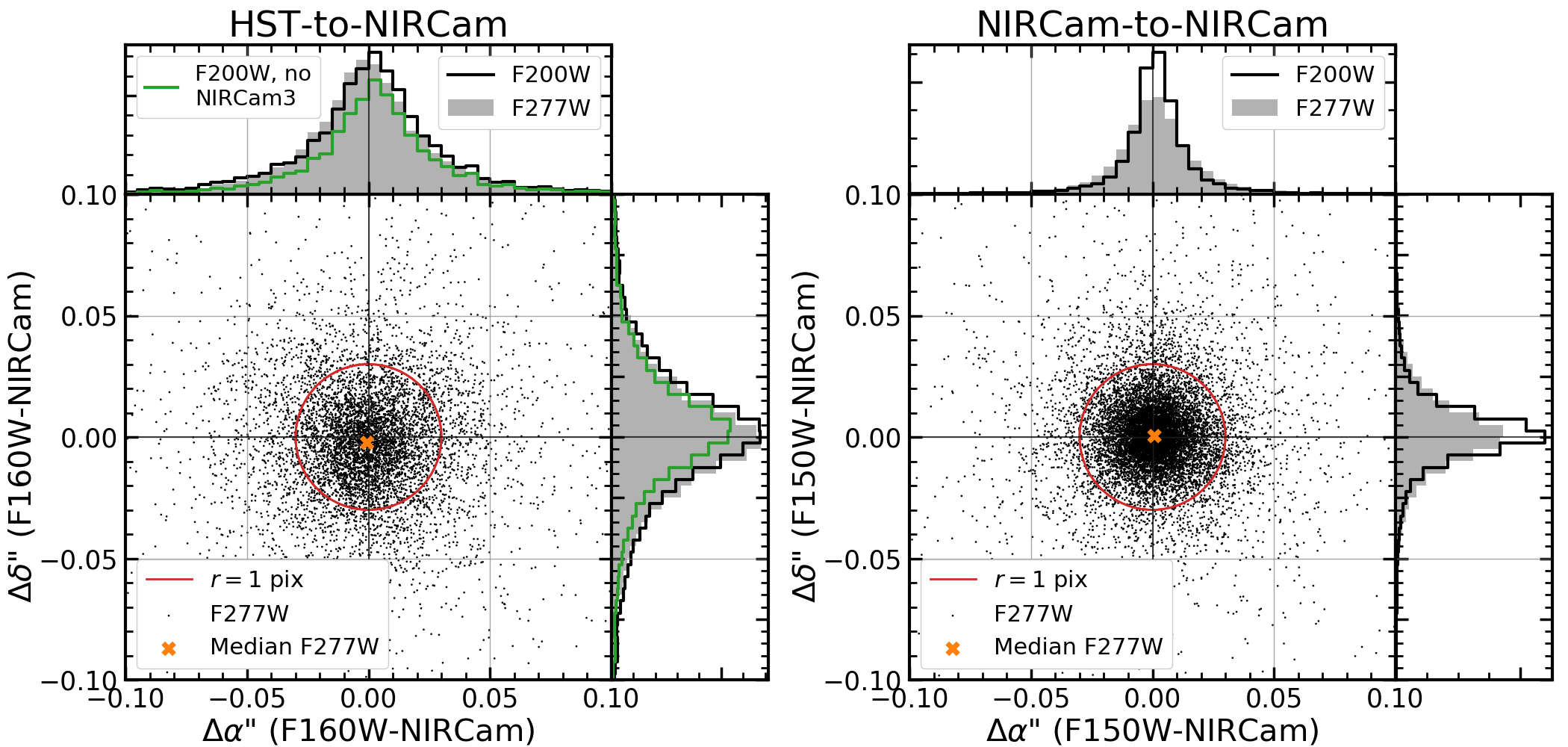}
\caption{Diagnostic plots of the CEERS NIRCam astrometric alignment. In the 
lefts panel we show the offsets for sources detected in all four pointings 
when compared with their positions in the \hst\ F160W mosaic. The right panels 
show astrometric offsets compared with positions in the NIRCam F150W mosaic. 
The black points in the scatter plots show the offsets for source positions 
in the F277W images, and the orange `x' indicates the median astrometric offset.
The filled histograms show the distribution of offsets in right ascension and
declination for F277W source positions, and the empty histogram shows the same
for source positions in the F200W images. We show both a SW and LW filter  
to indicate the quality of the astrometry in both channels. The red circles 
have a radius of 0\farcs03, or one pixel in the output mosaics. Both cases 
show a median offset consistent with 0, and an RMS scatter less than one
pixel. The RMS of the \hst-to-NIRCam alignment is $\sim12-15$ mas while that
of the NIRCam-to-NIRCam alignment is $\sim5-10$ mas, generally smaller for the SW and larger for the LW filters.
In the left panel, the green histogram indicates the F200W distribution excluding 
NIRCam3, the field that exhibits larger offsets relative to \hst\ (see the text for 
details). The F277W distribution excluding NIRCam3 is very similar.
\label{fig:astrometry}}
\end{figure*}

We perform an astrometric calibration using a modified version of the 
\pipeline\ \texttt{\textsc{TweakReg}} routine, with the modifications primarily aimed at exposing more fitting parameters and being able to input our own catalogs (similar modifications are also in progress in the official STScI pipeline). The \texttt{\textsc{TweakReg}} routine
calculates the transformation needed to align images to an absolute WCS frame.
It does this by comparing the source positions detected in each input image
with their matches in a reference catalog. With our modified version of this 
step, we create a \se\ \citep{bertin1996} catalog for each individual input 
image, providing improved source detection, deblending and centroiding over 
the internal \texttt{\textsc{TweakReg}} source identification.
We also use a custom reference catalog derived from a \hst\ F160W 
0\farcs03/pixel mosaic\footnote{\url{ceers.github.io/hdr1.html}\label{note:hdr1}} 
in the EGS field with astrometry tied to Gaia-EDR3 
\citep[see][for further details on the methods for image processing and mosaic 
creation]{koekemoer2011}. In both cases we use the \se\ windowed coordinates 
for improved centroiding, which especially improves positions for the compact sources that we primarily use for our alignment.

We calculate both a relative and absolute astrometric correction.
The images are first aligned relative to each other, where shifts in $x$ and 
$y$ are determined between images of the same detector. This step accounts
for uncertainties in the guide star alignment and pointing accuracy during
dithering. The RMS of this relative astrometry is $\sim3-6$ mas per source. The images 
are then aligned to the \hst\ F160W reference catalog, allowing for $xy$ 
shifts and rotations. In the LW images we also allow for a scaling factor to 
account for any additional distortion across the larger detectors, though we 
note that the 
calculated deviations from a unity scaling factor are small ($\sim1\times10^{-5}$). 
The RMS of this absolute alignment (WFC3-to-NIRCam) is $\sim12-15$ mas, and 
is driven by the larger F160W PSF. The RMS of the alignment between NIRCam 
images in different filters (NIRCam-to-NIRCam) is $\sim5-10$ mas, with 
SW images at the smaller end and LW images at the larger end of that range.
Figure~\ref{fig:astrometry} illustrates the quality of the astrometry in 
F200W and F277W and shows both the \hst-to-NIRCam and NIRCam-to-NIRCam 
alignment.

In summary, for all four NIRCam fields (1, 2, 3, and 6), for all filters and all detectors in both modules, the RMS astrometric alignment quality is generally $\sim5-10$ mas per source between NIRCam filters (SW - LW), and $\sim12-15$ mas per source relative to the absolute frame defined by the HST F160W mosaic. There is however one exception to this:  for NIRCam pointing 3, specifically for a $\sim1\arcmin$ region in the quadrant of the mosaic mostly covered by the B2 detector, the astrometric offsets relative to the HST F160W mosaic are larger, $\sim$0\farcs05 (or about one-fifth of the  0\farcs23 \hst\ F160W mosaic PSF in this mosaic\footnote{Note that, while the FWHM of the optical PSF of \hst\ is 0\farcs151 for F160W, this is subsequently convolved by: (1) the WFC3/IR detector pixel size when the field is imaged, (2) by a factor \texttt{pixfrac=0.8} times the WFC3/IR detector pixel size, and (3) by the mosaic pixel size, therefore yielding a final PSF FWHM of $\sim$0\farcs23 in the F160W mosaic, which is standard for this instrument.}).
These offsets are also present in this region of  the LW BLONG module, suggesting that it is not an alignment problem with a specific SW
detector. After detailed examination of the relevant datasets, a possibility is that the \hst\ F160W exposures for the particular visit in that part of the mosaic were affected by a low-level guidestar tracking issue, at the sub-PSF level ($\sim$1/5 of a PSF), which can occasionally occur and is not always revealed by the guidestar telemetry keywords in the image headers. Since these residuals are small compared to the size of the F160W PSF, they are unlikely to significantly affect the overall \hst-NIRCam photometry since our apertures are generally substantially larger than the \hst\ PSF.
We therefore followed a different process for aligning this NIRCam pointing. 
We first aligned F277W to F160W, excluding all sources in the upper left 
quadrant of detector BLONG from the fit. We then used F277W rather than 
F160W as the reference catalog for all other filters in this pointing. The 
RMS of the alignment is still $\sim 3-6$ mas (relative) and $\sim 5-10$ mas
(absolute NIRCam-to-F277W).

\begin{figure*}
\epsscale{1.0}
\plotone{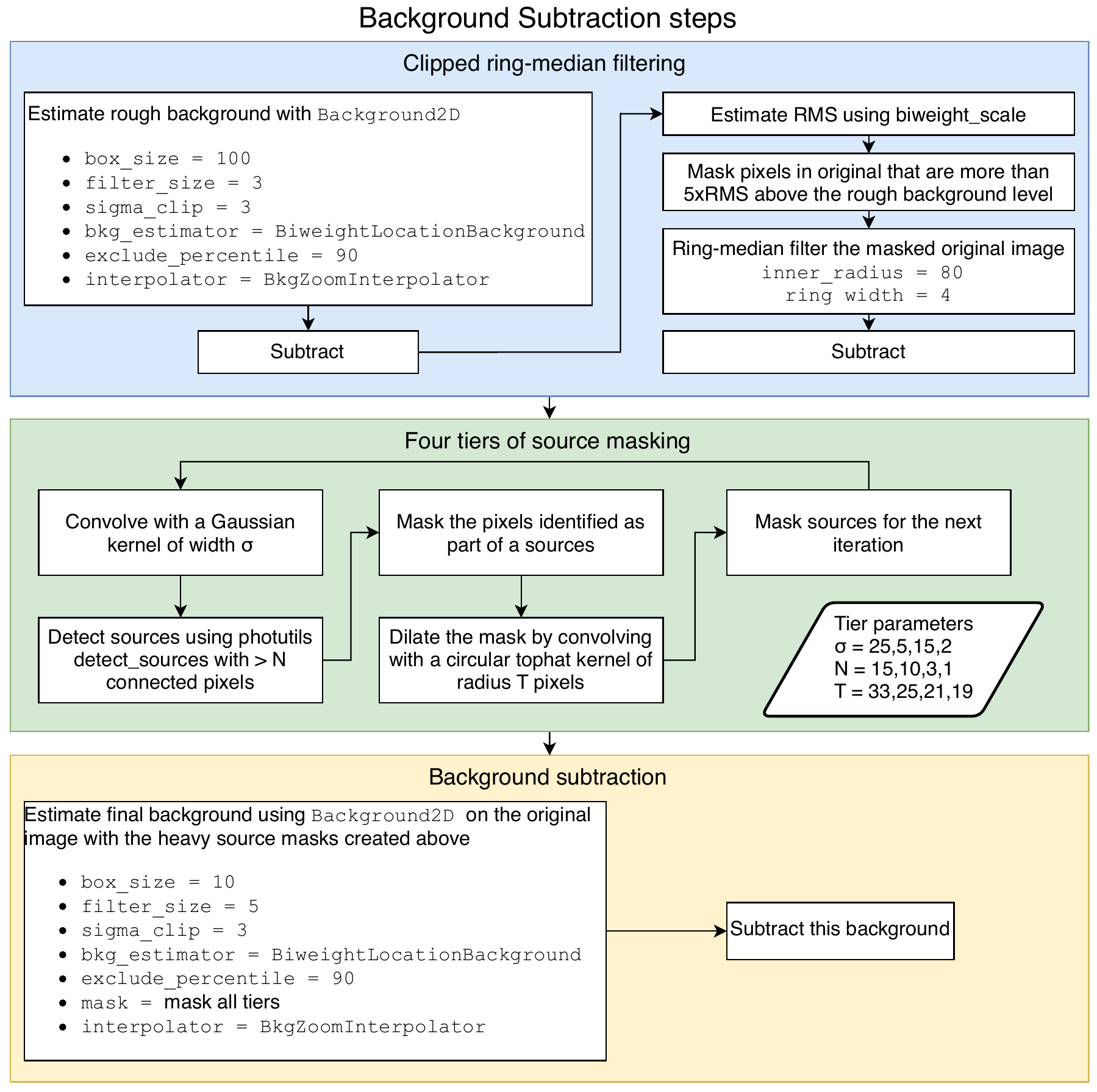}
\caption{Steps for subtracting background from the co-added images. 
\label{fig:background_steps}}
\end{figure*}

\subsubsection{Mosaic Creation} \label{sec:mosaics}
Before combining individual exposures into mosaics, we perform three additional
corrections to the calibrated images. First, we subtract a pedestal value in 
MJy/sr from each exposure. This step is necessary because the 
\texttt{SkyMatch} routine of \pipeline\ does not successfully match the 
background across all detectors, likely because the small dithers result in
little to no overlap between detectors. In order to calculate the background,
we mask bad pixels and source flux using the tiered masks created for each
image in Section~\ref{sec:1fnoise}. We then fit a Gaussian to the distribution
of unmasked, sigma-clipped pixel fluxes and take the position of the peak as 
the image pedestal. 
This single value is subtracted from each image, and 
the headers are updated to reflect this value.
Second, we calculate the sky variance in each background-subtracted image. To do this, we mask sources in four tiers, as described in Section \ref{sec:background}. We then block-sum the image in units of $7 \times 7$ pixels and use astropy's \texttt{biweight\_midvariance} to make a robust estimate of the variance in blocks that have had no pixels masked. We estimate the equivalent per-pixel variance by dividing by the number of pixels per block (49). We then scale  
the readnoise variance array
(\texttt{VAR\_RNOISE}) to reproduce this value. Because  \texttt{VAR\_RNOISE} is used to compute the inverse 
variance that is used as weighting during the drizzle process, this ensures that the resulting error arrays do a good job of predicting the RMS sky fluctuations (when estimated over large enough scales to avoid the pixel-to-pixel correlations introduced by drizzling).

Third, we make one final correction to the variance arrays to ensure that bad pixels are properly flagged. With our simulated mosaics (see Appendix~\ref{sec:simsmosaics}), we found that some known bad pixels had values of exactly zero in the variance arrays. When the dithered images were co-added, these areas in the output error array had relatively low RMS compared with the average. In these areas, the input error array with the missing data or bad pixel did not contribute to the RMS of the affected pixel. 
As a result, there were ``holes'' in the output error arrays in sets of three, matching our three dithers. Spurious source detections were more prevalent in these areas because the RMS used for detection was low. For each individual image, we therefore replace any pixels that are exactly zero in the three variance arrays with values of infinity (\texttt{numpy.inf}). This correction ensures that the affected pixels are correctly down-weighted in the drizzling. 

We create individual mosaics for each pointing using the Stage~3 routine
\texttt{Resample}, which uses the drizzle algorithm with an inverse variance 
map weighting \citep{fruchter2002,casertano2000} to combine images into a 
single distortion-free image. The output mosaics are drizzled onto a common 
WCS with the same tangent point as the \hst\ mosaics we have created in 
this field (see footnote~\ref{note:hdr1}). All \hst\ and NIRCam mosaics are
therefore pixel-aligned across all filters. The mosaics have pixel scales of 
0\farcs03/pixel, and we chose not to shrink the input NIRCam pixels when drizzling (i.e., \texttt{pixfrac}~$=1$). This larger \texttt{pixfrac} is 
preferred for CEERS because the majority of the mosaic is covered by at most
three exposures. The output pixel scale is only slightly smaller than the
original scale in the SW channel (0\farcs031/pixel) and just over two times
smaller than that of the LW channel (0\farcs063). We discuss tests to 
evaluate the optimal drizzle parameters in Appendix~\ref{sec:simsmosaics}. The choice of \texttt{pixfrac} leads to increased pixel correlations in the output image, which we discuss in Section~\ref{sec:issues} and have not yet addressed in our mosaics.

\begin{figure*}
\epsscale{1.15}
\plotone{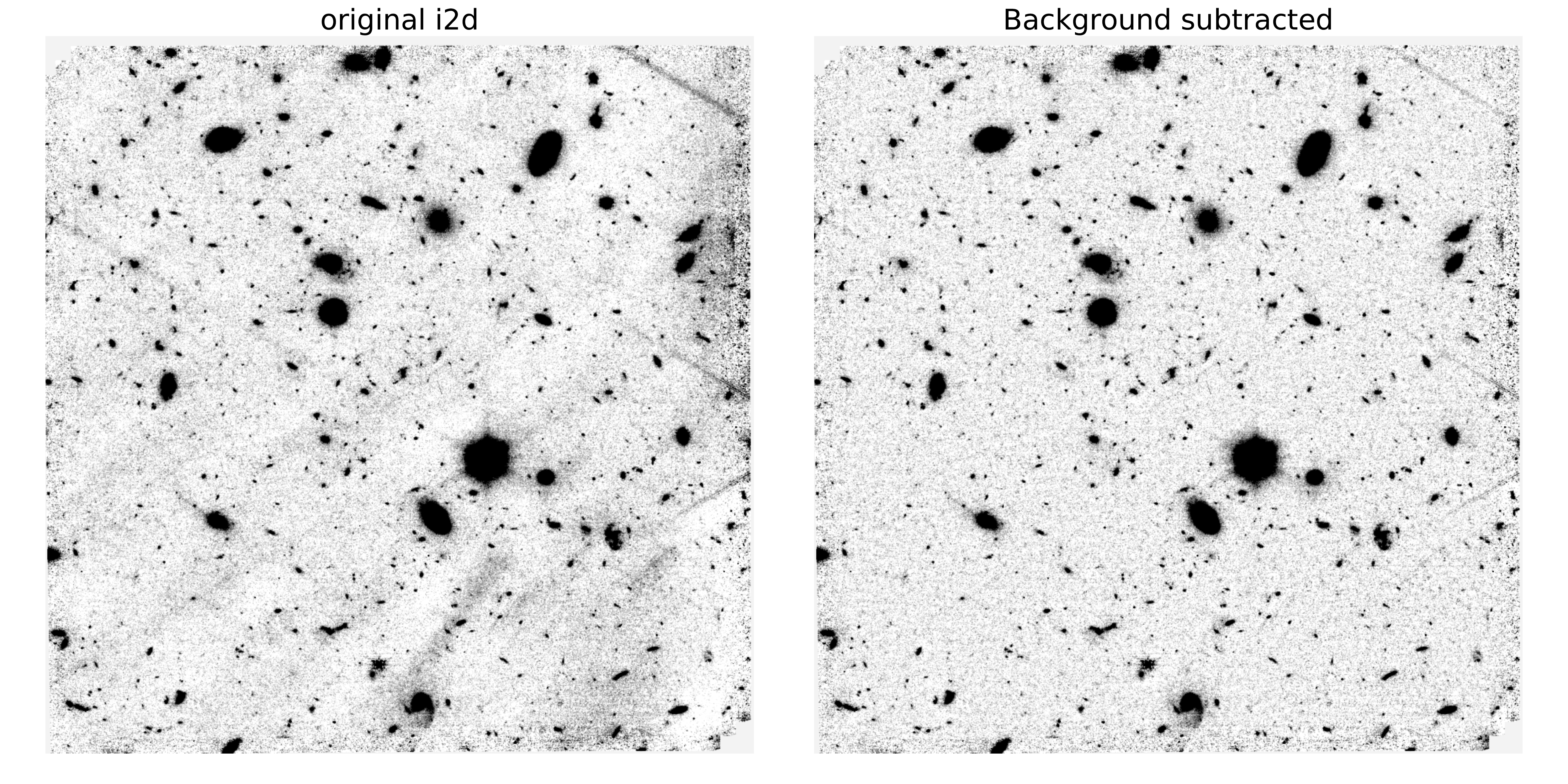}
\caption{Example of background removal for a single detector in the F200W of CEERS field 1. The image on the left is the co-added image before background subtraction.  The image on the right has had the fluctuating background removed as described in the text.
\label{fig:bkgsub_image}}
\end{figure*}

\begin{figure*}
\epsscale{1.15}
\plotone{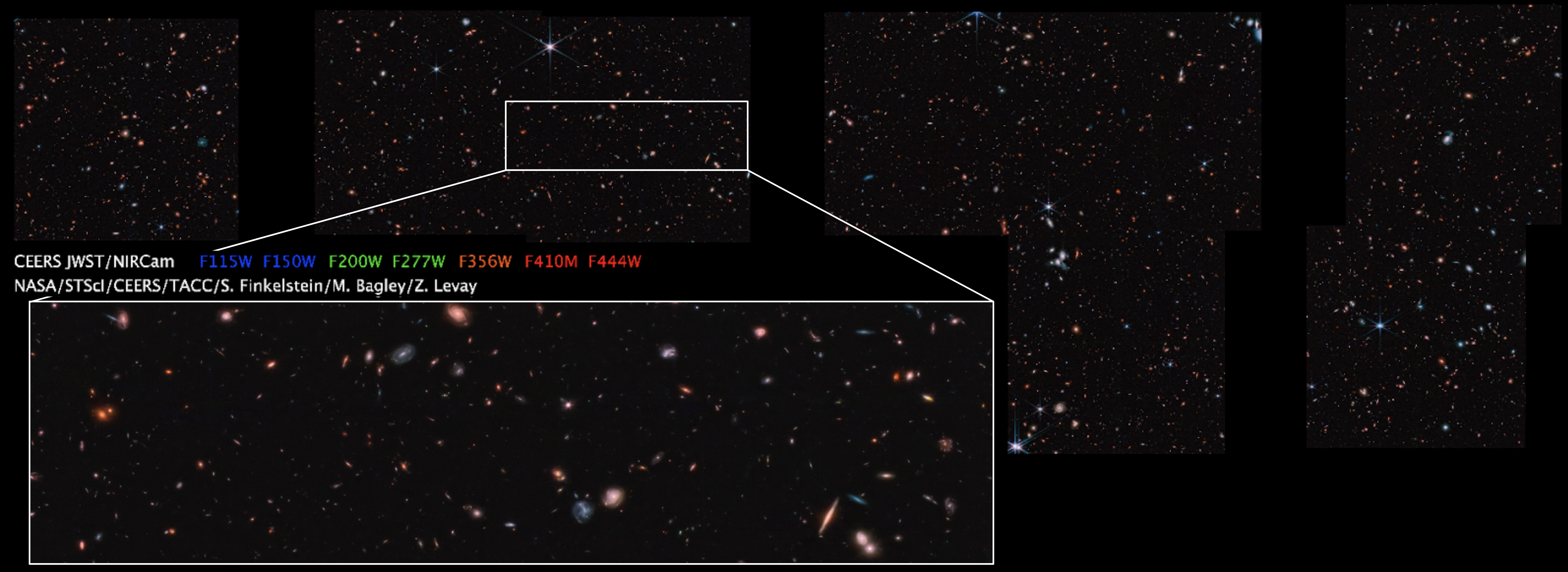}
\caption{A seven color image of the CEERS Epoch 1 NIRCam imaging. These images
were reduced at the Texas Advanced Computing Center and colorized by Zoltan Levay. High-resolution versions of these images and related information is available at \url{ceers.github.io/ceers-first-images-release.html}.
\label{fig:nircam_color}}
\end{figure*}

These individual mosaics are included in our data release and the files 
and extensions are described in Appendix~\ref{sec:files}. We also created 
a set of combined, Epoch 1 mosaics that include all four pointings. Constructing
these Epoch 1 mosaics requires considerable memory ($\sim$600 GB, depending on the filter), and 
so we use the Frontera computing system at the Texas Advanced Computing 
Center (TACC). Frontera provides a set of large memory nodes that offer 2.1 
TB of memory each with the option to link multiple nodes. A color image of 
the Epoch 1 mosaic including all seven filters is displayed in 
Figure~\ref{fig:nircam_color}.

\subsubsection{Background Subtraction} \label{sec:background}

Finally, we estimate and subtract any remaining background in the mosaics using 
a custom Python script that efficiently masks source flux before fitting
the unmasked pixels with a two-dimensional model. The steps are outlined in Figure \ref{fig:background_steps}. The first step involves taking out large-scale fluctuations, masking bright pixels in the residual, and then running a large ring-median filter across that image to create the first good estimate of the background. The inner ring radius is 2.4\arcsec\, so this step preserves the wings of all but the largest galaxies in the image. This step creates background that is flat enough on large scales to permit us to mask sources above a fixed threshold without any perceptible gradients in source size and density due to large-scale changes in the background. We mask four tiers of sources, with tier number 1 intended to mask the most extended galaxies, moving progressively to the smallest galaxies at tier number 4. This is achieved by convolving the image with four different widths of Gaussians, and masking areas with a minimum number $N$ of connected pixels above a fixed threshold of 1.5$\sigma$. These masks are then grown by dilating them with a circular tophat kernel with different kernel radii for each tier. We construct a source mask for each filter, including the pixel-aligned images in six available \hst\ filters (F606W, F814W, F105W, F125W, F140W and F160W). We merge all 13 separate masks into one,
which is used to mask source flux in all filters. In this way we can exclude
flux from sources that may be just below the detection threshold in some
filters but not others.

\begin{figure*}
\plotone{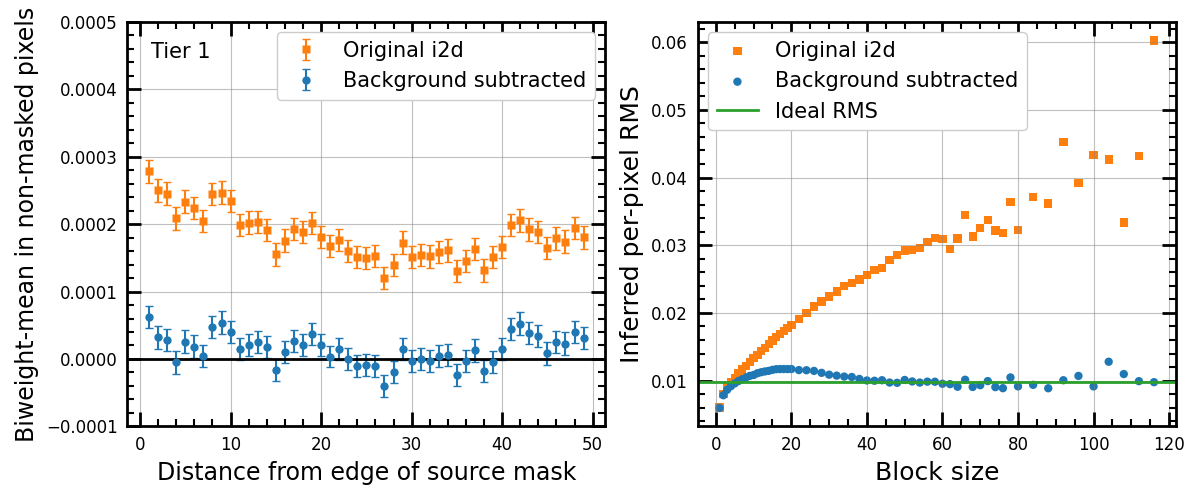}
\caption{
\textit{Left}: Sky background around the borders of tier 1 of the source mask. Tier 1 has the largest convolution kernel, and therefore masks the most extended sources. The diagram here shows the background (in MJy/sr) in unmasked pixels beyond the boundaries of the galaxies in this mask, as a function of distance to the closest masked pixel. One can see the overall elevated background level in the points from the original i2d file. There is a slight gradient toward brighter pixels closer to the boundaries of the masks, as expected from the faint extended wings of galaxies. The blue points show that the typical background is zero, and that wings of the galaxies have been slightly suppressed.
\textit{Right}: Statistics as a function of scale before and after background removal in the F200W image of NIRCam1. For each block size, the image has been block summed in boxes of $ N \times N $ pixels. The variance is then computed for blocks that had no pixels masked during the background estimation (i.e. these are our best effort at source-free regions of the image). The equivalent per-pixel RMS from this estimate is $\sqrt{var}/N$. The image co-addition suppresses the pixel-to-pixel RMS on scales of a few pixels. On larger scales, the equivalent per-pixel RMS continues to grow in the original image due to the various non-uniformities. Some of these non-uniformities are from scattered light, but the procedure also suppresses the extended wings of bright galaxies. The fluctuations are suppressed on scales of more than about 40 pixels by the background-subtraction procedure. The remaining fluctuations on intermediate scales are due to the wings of galaxies and to galaxies below the detection threshold. The green line shows the ideal RMS expected if the sky fluctuations were due purely to photon counting statistics. This helps to confirm that the procedure has done an excellent job of removing large-scale non-uniformities in the image. 
\label{fig:bkgsub_diagnostics}}
\end{figure*}

We then measure the background in the unmasked regions using the \photutils\
\texttt{Background2D} class. We use the \texttt{biweight\_location}  estimator 
to robustly calculate the average background in sigma-clipped boxes of 10$\times$10 
pixels in a grid across the image. The resulting low-resolution gridded 
background model is then median-filtered over 5$\times$5 adjacent boxes, and 
the \texttt{BkgZoomInterpolator} algorithm is used to interpolate the 
filtered array and construct a smooth background model.

A ``before and after'' example of background removal is shown in Figure \ref{fig:bkgsub_image}. 
Any estimation of background requires decisions about what is ``source'' and what is ``background.'' The procedure outlined above does a very good job of removing residual wisps and other artifacts in the NIRCam images, while preserving the wings of most of the galaxies of interest. However, it does suppress the wings of bright galaxies (intentionally) to enable the detection of faint neighbors.

To assess the background subtraction, we examined the stacked flux in pixels 
as a function of distance from the edge of the four separate tiers of masks. In most bands, for most tiers, there is very little ``spillage'' of galaxy 
flux into the pixels that were used to estimate the background. This spillage is impossible to avoid entirely. It is at a low enough level that we do not think the wings of galaxies that are above our detection threshold are significantly biasing the large-scale background estimation. In the left panel of Figure~\ref{fig:bkgsub_diagnostics} we show an example for F200W in one of our fields. 

On large scales, one would like to see the RMS of the background approach 
the RMS expected from a completely flat image affected only by the counting
statistics of the incoming signal. 
The right panel of Figure~\ref{fig:bkgsub_diagnostics} shows the per-pixel RMS that is inferred from measuring the RMS in source-free (i.e. unmasked) regions of a F200W image, when block summed over successively larger block sizes and then rescaled back to the per-pixel RMS. 
On small scales, the RMS is suppressed in the combined images because the drizzling procedure introduces a correlation between pixels. On medium scales, the residual wings of galaxies -- which are impossible to remove entirely -- slightly boost the RMS. The RMS approaches the ideal on larger scales, albeit with scatter in this measurement because there are very few blocks of $60 \times 60$ pixels and above that are entirely source free. 

\subsubsection{Field Depths} \label{sec:depths}
In Table~\ref{tab:depths}, we present the $5\sigma$ limiting magnitudes in each 
filter and pointing. These depths are measured  on our background-subtracted mosaics in circular apertures with radius
$r=0.1"$, and scaled to an empirical estimate of the noise in the image as described by Finkelstein et al. (2022, in prep). Briefly, apertures ranging in size from $r=0\farcs05$ to $r=1\farcs5$ are placed across each image, avoiding source flux and bad pixels. A robust estimate of the $1\sigma$ noise is calculated in each aperture, and following the examples of \citet{papovich2016}, \citet{bagley2022a} and \citet[][see also \citealt{labbe2003,blanc2008,whitaker2011}]{finkelstein2022a}, a second order polynomial is fit to the noise as a function of aperture size. We use this function to estimate the noise in a $r=0\farcs1$ aperture. Using stacked PSFs in each filter, we then measure the fraction of the total flux that is enclosed in an aperture of this size and correct the noise estimate by this amount. See Finkelstein et al. (2022, in prep) for a full description of each step in this process. The $5\sigma$ depths for each pointing and filter are listed in Table~\ref{tab:depths}.

\begin{deluxetable}{ccccc}
\tablecaption{NIRCam Pointing 5$\sigma$ Depths\label{tab:depths}}
\tabletypesize{\small}
\tablehead{
\colhead{Filter} & \colhead{NIRCam1} & \colhead{NIRCam2} 
& \colhead{NIRCam3} & \colhead{NIRCam6}}
\startdata
F115W & 29.08 & 29.10 & 29.21 & 29.21 \\
F150W & 28.96 & 28.94 & 29.05 & 29.04 \\
F200W & 29.17 & 29.16 & 29.21 & 29.16 \\
F277W & 29.16 & 29.20 & 29.22 & 29.19 \\
F356W & 29.14 & 29.17 & 29.18 & 29.18 \\
F410M & 28.37 & 28.35 & 28.41 & 28.40 \\
F444W & 28.57 & 28.58 & 28.58 & 28.58 \\
\hline
F606W & 28.62 & 28.62 & 28.62 & 28.62 \\
F814W & 28.30 & 28.30 & 28.30 & 28.30 \\
F105W & \nodata & 27.11 & 27.11 & 27.11 \\
F125W & 27.31 & 27.31 & 27.31 & 27.31 \\
F140W & 26.67 & 26.67 & 26.67 & 26.67 \\
F160W & 27.37 & 27.37 & 27.37 & 27.37 \\
\enddata
\tablecomments{Depths are measured for point sources in $r=0.1$\arcsec\ 
circular apertures and corrected to total flux using the flux measured
for a stacked PSF within that aperture.
The depths for NIRCam2 reported here are for the first set of 
observations only (obtained 21 June 2022) and do not include the 
additional imaging in F200W and F444W. All values are AB magnitudes.
The depths in the \hst\ filters are measured across the full EGS mosaics,
and so have the same reported values here for each individual NIRCam 
pointing. There is no WFC3 F105W coverage for NIRCam1.
}
\end{deluxetable}

\subsubsection{Flux Calibration} \label{sec:zeropoints}
As mentioned in Section~\ref{sec:stage2}, the CRDS pmap we use (0989)
for our NIRCam reduction includes updated photometric calibration reference 
files with zeropoints derived from early observations of three standard stars.
Due to the nature of the CEERS observing strategy, the NIRCam dithers are not large enough to observe a given source across multiple detectors, and so we do not have a direct test of detector-to-detector zeropoints. We therefore explore the relative photometric calibration between detectors with two methods, which we describe below. 

For the first method, we measured aperture
photometry on NIRCam, \hst/WFC3, and {\it Spitzer}/IRAC data for galaxies with $m <$ 24 using \cite{kron1980} apertures. We compared the NIRCam F115W filter to WFC3/F125W and F150W to F160W, after first matching the PSFs. We note that the passbands of these two pairs of filters are not exactly the same, preventing us from using this method to obtain an absolute photometric calibration. Moreover, the derivation of absolute zeropoints is challenging with extended sources due to uncertainties in aperture corrections, making it difficult to separate the color-dependent correction for missing source wings from the flux calibration.
Yet the approach of comparing NIRCam, WFC3 and IRAC fluxes in PSF-matched images should be robust for estimating relative calibrations between detectors provided enough sources are used. 

\begin{figure*}
\epsscale{1.0}
\plotone{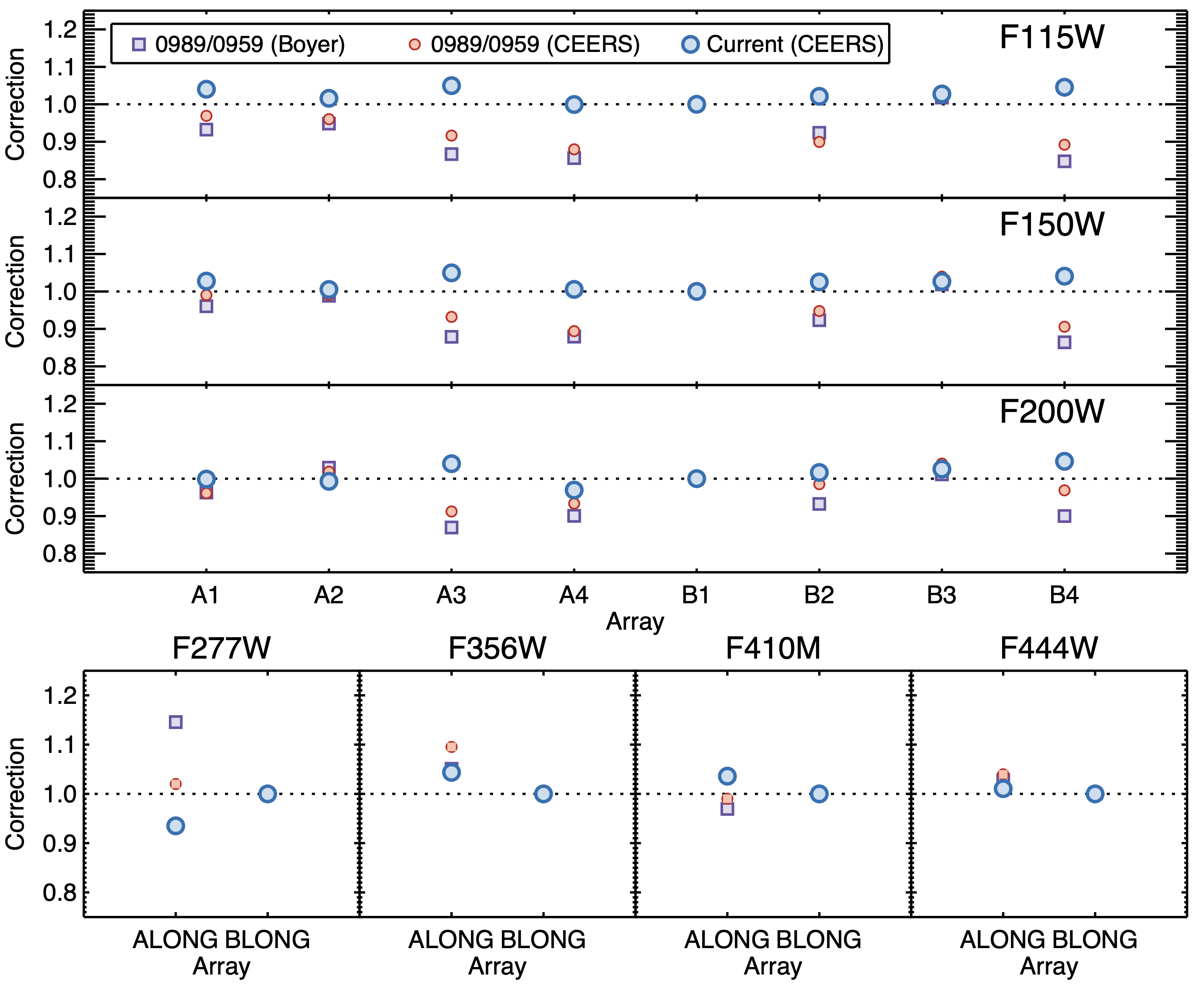}
\caption{Evaluation of the relative detector-to-detector photometric calibration for the CEERS NIRCam data. Squares show the relative multiplicative offsets with respect to the B1 (for SW) and B5 (for LW) detectors released in pmap-0989, which were calculated with data calibrated with pmap-0959 for an A-type star, a white dwarf, a G-type star, and verified with a globular cluster (see \citealt{boyer2022a}). Our own calculations of those offsets based on photometry for galaxies in our CEERS imaging data reduced with pmap-0956 (directly comparable with pmap-0959) are shown with smaller red circles, presenting an overall good agreement with the official recalibration released in pmap-0989. A re-evaluation of the zeropoint detector-to-detector offsets for the DR0.5 images released through this paper (calibrated with pmap-0989), is shown with larger blue circles, indicating residual offsets at the 2-4\% level.
\label{fig:phot_cal}}
\end{figure*}

We use a second method for F200W, F277W, and F410M, for which there are no direct imaging data with comparable depth and spatial resolution available. We used the CANDELS photometric catalog \citep{stefanon2017} to obtain synthetic magnitudes in the NIRCam bands by convolving the filter transmission curves with the stellar population synthesis models best-fitting the spectral energy distributions of bright (F160W~$<24$~mag) galaxies. We then compared to \citet{kron1980} aperture photometry in the NIRCam images. We confirmed the robustness of this method by using it on the four filters for which we had imaging data in similar filters, described in the previous paragraph. We obtained consistent results, compatible within 1-2\%, but with a 1.5$\times$ larger scatter.

Figure~\ref{fig:phot_cal} shows our results comparing the detector-to-detector photometric calibration for pmap 0956 (before the recalibration released in pmap 0989, which we use in this paper), compared with the recalibration work carried out with observations of three standard stars (PIDs 1536, 1537, and 1538), verified with data for a globular cluster (PID 1334) as described in \citet{boyer2022a}, and used in a new NIRCam zeropoints release in pmap 0989. We calculate the relative offsets of all SW detectors to B1. We apply a similar procedure to the F356W and F444W images, comparing them with IRAC 3.6\micron\ and 4.5\micron\ observations, respectively. We then calculate the relative offsets for these LW filters with respect to the BLONG detector. Our detector-to-detector relative calibrations based on galaxy photometry match (well within 2\% in most cases) the results obtained with stars (which are better suited for flux calibration).

The two methods were also applied to the final mosaics described in this paper, in order to identify possible residual offsets in the relative calibration of the different detectors that might still be present in pmap 0989. The results are also shown in Figure~\ref{fig:phot_cal}. We detect offsets between detectors at the $<5$\% level, with a median of 2.5\%. We do not apply these small offsets to our images at this time, but warn the reader that the relative and absolute calibration of NIRCam images is still uncertain at the $\sim5$\% level.

Finkelstein et al. (2022, in prep) have also explored the accuracy of the photometric calibration by exploring what average offsets in flux calibration would be needed to minimize the colors between observations and best-fitting spectral templates for a sample of several hundred spectroscopically confirmed galaxies.  They also find that the calibration is good to 1-2\% in most bands, and no worse than $\sim$5\% (in the bluest bands).  Again we note that these corrections were derived based on galaxies, and thus depend on aperture corrections.  Future photometric calibration updates based on stellar observations will reduce the amplitude of these corrections.

\section{Known Issues}\label{sec:issues}
This first public data release of CEERS NIRCam imaging represents a
best-effort reduction given the version of the \pipeline\ pipeline and
calibration reference files available at the time.
Several aspects of our reduction will be improved in future versions.
We briefly summarize the known issues regarding this current
version.

First, the set of calibration reference files (pmap 0989) still includes
some that were created pre-flight.
As discussed in Section~\ref{sec:stage2} and \ref{sec:zeropoints}, the flats
and flux calibration reference files are still preliminary. The current
NIRCam flats are ground flats that have been corrected to account for
illumination variations present in early observations. According to the
CRDS database (see footnote~\ref{note:crds}), the large-scale
variations in the flat fields have been corrected to $\sim1-2$\%, and
ongoing programs to monitor sky flats and measure illumination patterns
will continue to improve the flats. The photometric calibration reference
files are the result of significant early efforts
by the GTO and calibration teams, yet are still based on only a handful
of observations. In Section~\ref{sec:zeropoints}, the photometric
residuals from detector-to-detector are at the $\sim$2-5\% level.
The relative and absolute calibrations will continue to improve with more
observations and monitoring of photometric standards observed on each of the
ten NIRCam detectors. These programs (e.g., PIDs 1475, 1476, PI Boyer) and
others will deliver iterative improvements to the reference files
throughout Cycle 1.

Second, as mentioned in Section~\ref{sec:wisps}, the templates used for wisp
subtraction include some source ghosts. At the time of their creation, there
were not enough observations with sufficient dithers to successfully remove
all input sources from a combined stack. As a result, some small regions are
oversubtracted (at the $\sim3-4$\% level) in the CEERS images when we
subtract the scaled templates. The wisp templates will continue to improve
as more programs obtain NIRCam imaging, allowing for cleaner stacks that
capture the wisp patterns without contributions from sources.
In the meantime, our background subtraction (Section~\ref{sec:background}) does an excellent job of removing any residual
wisps or other background variations that may be introduce by our wisp
subtraction.

Third, we note that the NIRCam3 and NIRCam6 mosaics exhibit more cosmic
rays than the NIRCam1 and NIRCam2 images. This situation is expected given
that the exposures in pointings 3 and 6 (a single dithered set of exposures
per filter) are approximately two times longer than those in pointings 1 and 2
(where the depth is achieved in multiple sets of dithered exposures). In this
reduction version, we have used the same reduction parameters for all four
NIRCam pointings. In the future, we will tune the jump detection step of
Stage 1 and the outlier detection step of Stage 3 to ensure that cosmic rays
and outliers are appropriately identified and flagged in all exposures.

Fourth, we have not yet performed a careful check for persistence in the
majority of the NIRCam images. We have masked out the very strong persistence
in the additional F200W imaging in NIRCam2 (Section~\ref{sec:ceers2b}), and
that correction itself was an initial attempt that can be improved. In a
future version we plan to stack each dither in pixel space (i.e., not aligned
in WCS space) and identify sources that do not move on the detector from
exposure to exposure and also perhaps fade with time.
However,
this method will not be able to identify persistence affecting extended
sources that are larger than the dither size.
We expect this approach to work well in NIRCam1 and NIRCam2, which have
a minimum of six exposures in most filters, but three exposures per filter
should be sufficient to identify compact cases persistence that can masquerade
as high redshift galaxies.

Fifth, in creating the resampled mosaics, we do not make any correction for
correlated noise. As discussed in Section~\ref{sec:mosaics} and
Appendix~\ref{sec:simsmosaics}, we use \texttt{pixfrac}~$=1$ so that each
output pixel is sufficiently covered by input pixels in pointings where there
are only three dithers. However, drizzling the full input pixels (i.e.,
without ``shrinking'' the pixels first) results in a mosaic with correlated
pixels. In this case, the noise in the resampled science images and variance
maps is reduced by the correlations, and so photometric uncertainties will
by underestimated \citep[e.g.,][]{casertano2000,labbe2003,blanc2008,whitaker2011,papovich2016}.
We have rescaled the readnoise variance map to account for the sky variance
(Section~\ref{sec:mosaics}), but we have not yet applied any scaling to
account for the pixel-to-pixel correlation introduced by the drizzling.
This scaling can be estimated by computing the autocorrelation function of
background, non-source pixels and determining the amount to which the noise
has been suppressed \citep[e.g.,][]{guo2013}.

Finally, we note that the F115W images are shallower than expected compared
with the pre-launch exposure time calculator. As the background at these
wavelengths is low, the images are dominated by readnoise to a larger degree
than expected. Updates to the reference files related to detector noise
properties as well as to processing methods may improve the depth in F115W
slightly. However, we expect that the CEERS F115W images will remain $\sim$0.2-0.3 magnitudes shallower than expected.

\section{Summary}\label{sec:summary}
We announce the public release of CEERS NIRCam mosaics covering four pointings 
in the EGS field. These images were observed in June, 2022, and obtained in 
parallel to primary MIRI imaging. The NIRCam imaging includes seven filters 
per pointing (F115W, F150W, F200W, F277W, F356W, F410M and F444W), reaching 
$5\sigma$ depths of $\sim 28.5 - 29.2$ for point sources.

We have reduced the images using version 1.7.2 of the \jwst\ Calibration 
Pipeline, with additional steps and modifications developed to handle 
challenges encountered with the data.
In Section~\ref{sec:realreduction}, we describe our image reduction process 
in detail, including our custom corrections for snowballs, wisps, $1/f$ noise, 
astrometric alignment, variance map scaling, and background subtraction. 
We produce mosaics in each filter and field, resampled onto a common output 
grid with a pixel scale of 0\farcs03/pixel and aligned with available \hst\ 
imaging in the field. 
Our reduction scripts and pipeline parameter files are summarized in 
Table~\ref{tab:reduction} and will be available and documented on GitHub. 
The images are the result of a best-effort reduction with the available 
\pipeline\ pipeline and calibration reference files early in Cycle 1. 
In Section~\ref{sec:issues}, we summarize issues with the current reduction 
version that will improve with updated information about instrument performance.

Our publicly released mosaics are available at \url{ceers.github.io/dr05.html} and 
on MAST via \dataset[10.17909/z7p0-8481]{\doi{10.17909/z7p0-8481}}. 
In the Appendix, we describe the file structure of the released images. We 
also describe previous CEERS NIRCam reductions that were used for results in 
some early CEERS publications. 

Our reduction process has been heavily influenced by our work with our pre-flight 
NIRCam simulations. We describe the creation of these simulations in 
Appendix~\ref{sec:sims}, using as input a mock galaxy catalog constructed with 
the Santa Cruz semi-analytic model, point sources and real 
$z\sim9$ candidate galaxies. We describe the simulated image reduction 
in Appendix~\ref{sec:simsreduction}, including a 
test of drizzle parameters that informed our parameter choices with the 
real images. Our NIRCam simulations are available for download at 
\url{ceers.github.io/sdr3.html}.

\begin{acknowledgements}
We acknowledge that the location where this work took place, the University 
of Texas at Austin, that sits on indigenous land. The Tonkawa lived in 
central Texas and the Comanche and Apache moved through this area. We
pay our respects to all the American Indian and Indigenous Peoples and 
communities who have been or have become a part of these lands and territories 
in Texas, on this piece of Turtle Island.

We thank Nikko Cleri, Rosemary Coogan, Asantha Cooray, Carlos G{\'o}mez-Guijarro, Nimish Hathi, Benne Holwerda and Marc Huertas-Company for a careful read of the manuscript.
We thank the entire \jwst\ team, including the engineers for making possible 
this wonderful over-performing telescope, the commissioning team for obtaining 
these early data, and the pipeline teams for their work over the years 
building and supporting the pipeline. 
The authors acknowledge the Texas Advanced Computing Center (TACC) at The 
University of Texas at Austin for providing HPC and visualization resources 
that have contributed to the research results reported within this paper. 
This work is based on observations with the NASA/ESA/CSA James Webb Space Telescope obtained from the Mikulski Archive for Space Telescopes at the Space Telescope Science Institute (STScI), which is operated by the Association of
Universities for Research in Astronomy (AURA), Incorporated, under NASA contract NAS5-03127.
We acknowledge support from NASA through STScI ERS award JWST-ERS-1345. 
We thank Zolt Levay for making the beautiful color images of the CEERS NIRCam
and MIRI observations. 

In this paper we refer to the \textit{James Webb Space Telescope}
using only the acronym \jwst,
reflecting our choice to celebrate the promise of this telescope without
acknowledging the public official for whom it is named.
This individual has been implicated in anti-LGBTQI+ attitudes that do
not reflect the authors' values related to inclusion in science.

\end{acknowledgements}

\vspace{5mm}
\facilities{\jwst\ (NIRCam), \hst\ (ACS, WFC3), \spitzer\ (IRAC)}

\software{Astropy \citep{astropy},
          Besan\c{c}on Model of the Galaxy (\url{doi:10.25666/osu-theta.20210107.galaxy-model}),
          Drizzle \citep{fruchter2002},
          \eazy\ \citep{brammer2008},
          \mirage\ (\url{mirage-data-simulator.readthedocs.io}),
          \photutils\ \citep{bradley2020},
          SciPy \citep{virtanen2020},
          \se\ \citep{bertin1996},
          STScI \jwst\ Calibration Pipeline (\url{jwst-pipeline.readthedocs.io})}

\appendix

\section{Mosaic Image Structure} \label{sec:files}
We have produced a multi-extension fits file for each of the seven NIRCam 
filters for CEERS NIRCam pointings 1, 2, 3 and 6. These files are the 
result of all reduction steps described in Section~\ref{sec:realreduction}.
The files are available and documented at \url{ceers.github.io/dr05.html}
and at MAST via \dataset[10.17909/z7p0-8481]{\doi{10.17909/z7p0-8481}}. The mosaics have 12 extensions,
which we briefly summarize here:
\begin{itemize}
\item \texttt{SCI\_BKSUB}: resampled, \textbf{background-subtracted} science data, in units of MJy/sr
\item \texttt{SCI}: resampled science data (not background subtracted), in units of MJy/sr
\item \texttt{ERR}: resampled uncertainty estimates as standard deviation, constructed as the sum in quadrature of the resampled variance maps
\item \texttt{CON}: context image, which encodes information about the input images 
    that contribute to each output pixel
\item \texttt{WHT}: weight image giving the relative weight of the output pixels, constructed from the \texttt{VAR\_RNOISE} array during resampling
\item \texttt{VAR\_POISSON}: resampled Poisson variance map
\item \texttt{VAR\_RNOISE}: resampled read noise variance map, which has been 
    rescaled to include a robust estimate of the sky variance
\item \texttt{VAR\_FLAT}: resampled flat-field variance
\item \texttt{BKGD}: the background model that was subtracted from the science image
\item \texttt{BKGMASK}: the tiered source mask used to create the background
\item \texttt{HDRTAB}: a table of FITS keyword values for all of the input 
    images that were combined to produce the output image
\item \texttt{ASDF}: metadata for the JWST data model
\end{itemize}
The extension \texttt{BKGD} can be computed by subtracting the \texttt{SCI\_BKSUB} extension 
from the \texttt{SCI} extension, but we include it in the mosaics for completeness.

We also provide images in six \hst\ ACS and WFC3 filters, cut out from the 
full EGS mosaics and pixel-aligned to each CEERS NIRCam pointing. 
While the background in the \hst\ filters was already very close to zero, we 
have background-subtracted these mosaics for consistency with the NIRCam images 
and provide the \hst\ mosaics both with and without this subtraction. 
The \hst\ mosaics have five extensions, which we summarize here:
\begin{itemize}
\item \texttt{SCI\_BKSUB}: resampled, \textbf{background-subtracted} science data, in units of counts/s
\item \texttt{SCI}: resampled science data (not background subtracted), in units of counts/s
\item \texttt{rms}: resampled uncertainty estimates as standard deviation
\item \texttt{BKGD}: the background model that was subtracted from the science image
\item \texttt{BKGMASK}: the tiered source mask used to create the background
\end{itemize}

\section{Earlier CEERS Data Reductions}\label{sec:prevreduc}
Here we briefly describe the previous versions of the CEERS NIRCam data 
reduction that were used for results presented in early papers from the 
CEERS team.

\subsection{Version 0.05, \citet{zavala2022a}}
\citet{zavala2022a} presented results based on version 0.05 of the CEERS
NIRCam reduction, which used \pipeline\ version 1.5.3 and pmap 0932 (NIRCam
imap 0214). This reduction involved a preliminary correction for $1/f$ noise, 
but did not correct for snowballs or wisps. The astrometric alignment for this 
version should also be considered preliminary. 
We calculated astrometric corrections for groups of images, rather than 
individually for each detector and dither. Specifically, we fit module A 
and B separately for the LW images (grouping all dithers for a given filter
and module together), and fit the SW images in three sets: (1) all module A
detectors, (2) detector B2, which demonstrated significant offsets from the 
other B detectors, and (3) the remaining three detectors of Module B. 
We first aligned the F200W images to the \hst/WFC3 F160W reference catalog 
tied to Gaia-EDR3, and then used \photutils\ to create a new reference catalog
in F200W. We cleaned the F200W reference catalog of all sources near detector 
edges and spurious sources around diffraction spikes and in the noise around 
bright sources, and considered only compact sources in the magnitude range
$18 < m_{200} < 27$. We aligned each NIRCam filter to F200w using the three
groups mentioned above. The median astrometric offset in each filter and 
NIRCam pointing is $\lesssim$0\farcs005, and the RMS is 
$\sim$0\farcs025$-$0\farcs03 ($\sim$1 pixel). However, for NIRCam pointing 2 
in particular, the astrometry in the F115W and F150W filter suffered from a 
larger global offset of $\sim$2-3 pixels. Additionally, the alignment between 
the images in each dither for this pointing was off by $\sim0.5$ pixels, which 
slightly smeared out a small percentage of source flux, especially problematic 
for compact sources.

\subsection{Version 0.07, \citet{finkelstein2022b}, \citet{guo2022} and \citet{kocevski2022a}}

\citet{finkelstein2022b}, \citet{guo2022} and \citet{kocevski2022a} presented results based on 
version 0.07 of the CEERS NIRCam reduction, which used \pipeline\ version 
1.6.2 and pmap 0942 (NIRCam imap 0221). This reduction version included 
wisp subtraction and an improved $1/f$ noise removal. It did not include a 
correction for snowballs, though we note that the authors of each paper carefully inspected 
individual exposures to ensure their results were not affected by snowballs. 
This reduction version also implemented the local background subtraction 
on the mosaics that is described in Section~\ref{sec:background}.

\citet{finkelstein2022b} presented an earlier arXiv version that used version
0.05 (see previous section), where the improved astrometric alignment had a 
significant effect on their results. The measured F150W flux of the 
high-redshift candidate Maisie's Galaxy increased sufficiently in v0.07 to change the best-fit 
photometric redshift from $z\sim14$ to a more tightly-constrained 
$z\sim12$. The authors have provided a full description of how the reduction version
changed their results.\footnote{\url{web.corral.tacc.utexas.edu/ceersdata/papers/Maisie_update.pdf}}

\section{Simulated CEERS NIRCam Imaging}\label{sec:sims}
In preparation for CEERS \jwst/NIRCam imaging data, we created a series of
simulated NIRCam images designed to reproduce the CEERS observing strategy. 
These simulations were designed to aid in our preparations with reducing the
real imaging as well as to validate our photometry and analysis methods for 
the real NIRCam imaging. The simulations are performed with the 
Multi-Instrument Ramp 
Generator\footnote{\url{mirage-data-simulator.readthedocs.io}} 
(\mirage), with input sources taken from a mock catalog created using the 
Santa Cruz semi-analytic model for galaxy formation. 
In the following sections, we describe the simulation inputs and creation.

\subsection{Santa Cruz SAM Mock Galaxy Catalog}
The mock images are simulated based on an augmented version\footnote{\url{https://ceers.github.io/sdr3.html\#catalogs}, SDR V3} of the mock galaxy catalog presented in \citet{yung2022a}. The simulated lightcone spans 782 arcmin$^2$ with coordinates overlapping with the observed EGS field and containing galaxies over redshift range $0 < z \lesssim 10$ and rest-frame $M_{\rm UV}$ range $-16\gtrsim M_{\rm UV}\gtrsim-22$. The galaxies in the mock lightcone are simulated with the Santa Cruz semi-analytic model (SAM) for galaxy formation \citep{somerville2015b, somerville2021}. 
Within dark matter halos extracted from the Bolshoi-Planck simulation \citep{klypin2016} and Monte Carlo merger trees generated with the extended Press-Schechter formalism \citep[e.g.][]{somerville1999a}, the SAM tracks the evolution of global galaxy properties under the influence of a set of carefully curated physical processes, including cosmological accretion, cooling, star formation, chemical enrichment, and stellar and AGN feedback. We refer the reader to \citet{yung2022a} for a schematic flowchart that illustrates the internal workflow of the SAM.
The model configuration and physical parameters are based on the calibration from \citet{yung2019a} and \citet{Yung2021}.
The models have been shown to reproduce the observed evolution in high-redshift (e.g. $z\gtrsim4$) one-point distribution functions of $M_\text{UV}$, $M_*$, and SFR \citep{yung2019a, yung2019b}, observational constraints on the IGM reionization history \citep{yung2020a, yung2020b}, as well as two-point auto-correlation functions from $0 \lesssim z \lesssim 7.5$ \citep{yung2022a}. 

The full star formation and chemical enrichment histories of individual predicted galaxies are forward modelled into rest- and observed-frame photometry with SEDs generated based on stellar the population synthesis (SPS) models of \citet{bruzual2003}. The observed-frame photometry is calculated accounting for dust attenuation effects using the attenuation curve of \citet{calzetti2000}, and for absorption by hydrogen along the line of sight in the IGM \citep{madau1996}.
In the CEERS mock catalogs, we also include self-consistently predicted nebular emission lines excited by young stars, AGN, and post-AGB stellar populations in the high-resolution synthetic spectra and the broad- and medium-band photometry (\citealt{hirschmann2017}, \citealt{hirschmann2019}; Yung, Hirschmann, Somerville et al., in prep.)
Galaxies are added to the images as \sersic\ profiles \citep{sersic1963,sersic1968}, with \sersic\ indices and effective radii determined as described in \citet{brennan2015}.

\subsection{Additional Simulation Inputs} \label{sec:simsinputs}
In addition to the mock galaxy catalog, we have added two additional sets of 
sources to the CEERS simulations. First, we have included the seven $z\sim9$ 
galaxy candidates in the EGS field presented by \citet{finkelstein2022a}. 
This sample represents a significant ($2\sigma$) overdensity in the EGS field
\citep{finkelstein2022a}, including two spectroscopically-confirmed galaxies at 
$z=8.683$ \citep{zitrin2015} and $z=8.665$ \citep{larson2022}. 
The seven galaxies are added as S\`ersic profiles at their expected coordinates,
and their \jwst\ photometry is estimated using the best-fit \eazy\ template
for each source's \hst\ and \spitzer\ photometry.

Second, we included point sources in the NIRCam imaging simulations in two 
ways. The positions and magnitudes of bright ($V<16$, Vega magnitudes) point 
sources in the EGS field are added from 2MASS, \textit{WISE} and \textit{Gaia}.
We added fainter point sources ($16 < V < 29$ Vega) using the Besan\c{c}on 
Model\footnote{\url{doi:10.25666/osu-theta.20210107.galaxy-model}} at mock 
positions to approximate the expected stellar density and luminosity 
distribution in the EGS field. In both cases, \mirage\ adds point sources
from a library of PSFs, which was constructed 
using \texttt{WebbPSF} \citep{perrin2014}. We use the default PSF library
included with the \mirage\ reference files. This gridded library includes an
image of the PSF core and an additional image of the PSF wings for each filter 
and pupil combination at a grid of positions across each detector.
For each source, the appropriate PSF is selected from this library, the core 
is combined with the image of the wings, and the full PSF is normalized and 
scaled to the required brightness. The separation of the core and wings allows
the user to tune the size of the PSFs in the simulated images. In the interest 
of decreasing processing time for the CEERS simulations, we adopted the 
default size of the PSF wing images (301$\times$301 pixels) at all magnitudes.
As a result, we note that the PSFs of some sources in our simulated images 
appear truncated. We found this an acceptable compromise as our science goals
for the simulations were focused on the identification and recovery of 
galaxy colors at high redshift.

\subsection{Simulating Raw Images with \mirage}
With these inputs, we use \mirage\ to construct raw NIRCam exposures that 
approximate the CEERS observations and include realistic estimates of noise, 
sky background level, and detector artifacts.
The Astronomer's Proposal Tool (APT) file defines the observation 
specifications, setting the filters, readout modes and patterns, and dithers.
At the time of our simulation creation, \mirage\ could not 
parse an observing setup with either MIRI or NIRSpec as the primary instrument 
and NIRCam imaging in parallel. However, as the parallel dithers are specified 
according to the primary observation (determined by either the size of the MIRI
PSF or the NIRSpec nodding pattern), the combined primary$+$NIRCam parallel 
specifications were crucial. We therefore created a separate, NIRCam-only APT 
file with the same targets, number of dithers, and the same exposure 
specifications as the real CEERS program. We then replaced the pointing 
information of this NIRCam-only mock program with that from the original CEERS 
APT file. In this way, we were able to create a set of parameter files as 
input to \mirage\ that replicated all of the CEERS dither patterns.
We note that this approach is no longer required, as \mirage\ can parse 
all observation templates and primary-parallel combinations.

The full CEERS NIRCam imaging simulation involves creating 1816 separate 
images, one image per exposure for each of the ten NIRCam detectors. This 
total includes 75 exposures obtained in parallel to NIRSpec MSA observations,
57 exposures in parallel to MIRI imaging, and 56 exposures taken with the 
NIRCam WFSS observing mode. For these simulations we include just the 
imaging portion of the NIRCam WFSS observations, which amounts to a total of 
24 direct image exposures in F356W and 56 SW exposures in F115W. 
We separately produced simulated CEERS NIRCam WFSS observations, which are 
available for download at \url{ceers.github.io/sdr3.html}.
  
For each simulated image, \mirage\ works in three stages. It first creates a 
noisless seed image of the input point sources and galaxies as well as any 
requested background signals. We adopted the beginning of the CEERS June 
observing window (15-17 June, 2022) as simulated observing date,
and \mirage\ used the \texttt{jwst\_backgrounds}\footnote{\url{github.com/spacetelescope/jwst_backgrounds}} tool to estimate the appropriate 
value for the EGS field on this date. In the second stage, \mirage\ 
reconstructs a dark exposure to match the required readout pattern with the
specified number of groups and integrations. The use of ground-tested dark 
exposures introduces real detector effects such as the bias, known hot pixels,
and detector noise properties. Finally, \mirage\ combines the dark 
with the seed image and adds additional effects such as cosmic rays.
We used \mirage\ version 2.2.1 and the CRDS pmap 0834 (NIRCam instrument map 0193) 
for the input reference files. 

\section{Reducing Raw Simulated Data} \label{sec:simsreduction}
While we leave the bulk of our discussion of our data reduction to 
Section~\ref{sec:realreduction}, here we summarize our steps 
processing the raw simulated data.
We do this for two reasons: (1) these processing steps informed our methods 
for working with the real CEERS NIRCam imaging, and 
(2) we aim to provide a reference for how the simulated products were 
processed for any who wish to use them for science.
 
We process the raw simulated images through the three stages of the \pipeline\
pipeline, with custom scripts we developed to handle differences between the
\jwst\ reference files available at the time of simulation and the input
assumptions of \mirage. The reduction uses \pipeline\ version 1.4.6 except 
where noted, and CRDS pmap 0834 for consistency with the simulated inputs. 
We adopt most of the default parameters for the \pipeline\ steps and describe
any changes below. For the detector-level corrections applied in Stage 1, we
apply the correction for interpixel capacitance, skipped by default, because
\mirage\ adds this effect to the simulated data. We also supply custom gain
maps for the steps that identify jumps in the up-the-ramp signal and fit the
ramp to obtain a countrate map. These custom gain maps were created to match
the average value per detector that \mirage\ uses in creating the images,
rather than the location-varying gain maps that were present in the reference
files at the time. 

Our first custom step is to remove the $1/f$ noise that is present in the 
images as horizontal and vertical striping. In short, this correction involves
masking out all bad pixels and source flux (using the seed images for source 
positions and footprints) and collapsing each image along columns and rows 
to measure the striping pattern. We developed this correction in working with 
this simulated data, and improved on the process for the real images. 
We describe this step in detail in Section~\ref{sec:1fnoise}. 

\begin{figure}
\epsscale{1.1}
\plotone{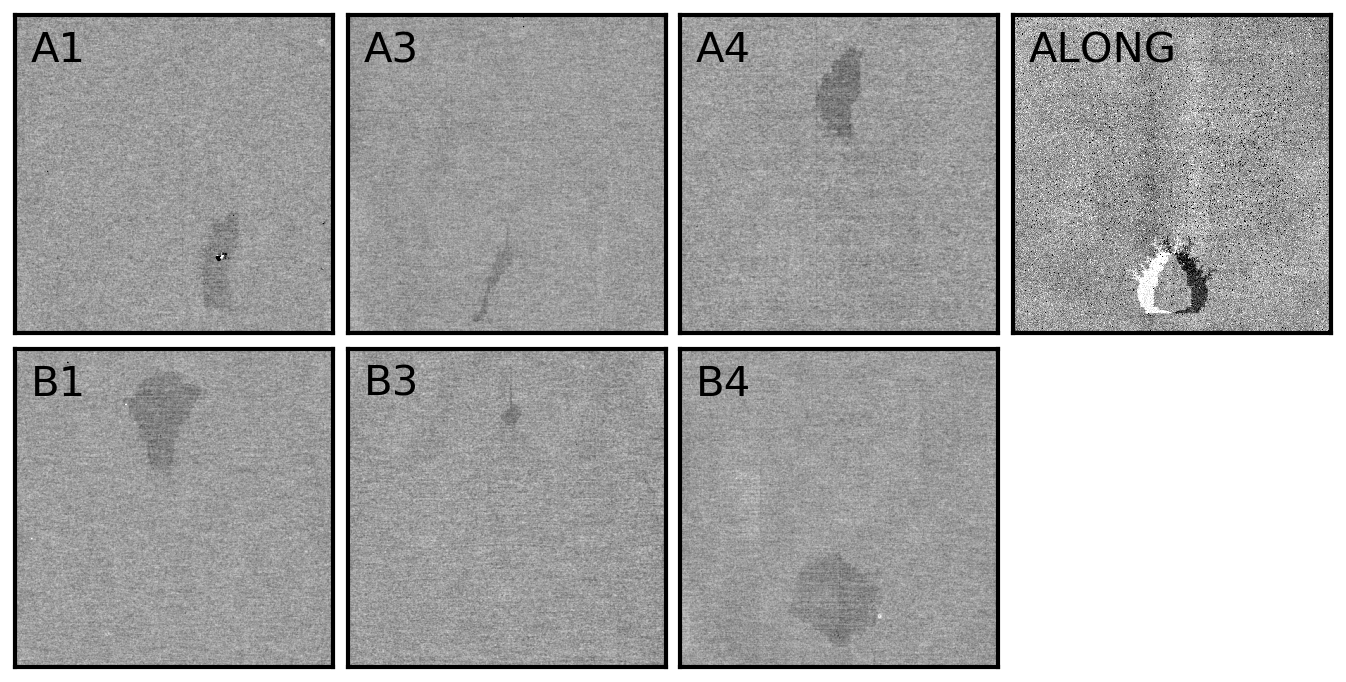}
\caption{Custom flat field images we created for the reduction of our 
simulated images. We show the SW detectors A1, A3, A4, B1, B3, B4, each of 
which has a different low-level feature present, caused by areas on the detectors
with thinner epoxy layers. We use these six custom 
flats in reducing our simulated images. We also show an earlier custom flat
we created for the LW detector ALONG. This feature was present in our 
SDR1 images and is included here for reference. With updated \mirage, 
\pipeline, and CRDS reference file versions, we no longer see this ALONG 
feature in our simulated images.
\label{fig:flats}
}
\end{figure}
\subsection{Custom Flat Fields} \label{sec:simscustomflats}
Following the calibration steps of Stage 2, we perform a correction to 
remove residual large-scale but low-level features present in some of the 
images. The features correspond to areas on the detectors where the epoxy 
layer is thinner than for the unaffected pixels.\footnote{e.g., STScI 
Technical Document \href{https://www.stsci.edu/files/live/sites/www/files/home/jwst/documentation/technical-documents/_documents/JWST-STScI-004622.pdf}{JWST-STScI-004622}, Hilbert \& Rest, 2014.}
We found that these epoxy voids are multiplicative features in our simulated 
images and can be removed in a similar fashion as applying a flat field.
The affected detectors are SW A1, A3, A4, B1, B3 and B4. In a previous version 
of our simulations (created using \mirage\ version 2.1.0, \pipeline\ version 1.3.3 
and pmap 0764), we found a prominent feature only on detector A5 that is not 
present in our new CEERS simulations. We believe the changing nature of these
epoxy voids in our simulations may be caused by updates in the reference files
used to add detector features to the images, such that the
presence and strength of the features is determined by a combination of CRDS 
context and \mirage\ and \pipeline\ versions. 

We therefore created custom flat fields for the six affected SW detectors in 
the following way. For each detector and filter we created 33 blank simulated 
images (chosen because each affected filter has at least 33 exposures 
across the full CEERS field) with no input sources and no cosmic rays added. 
We used the same input parameter files as used for the simulated images to 
ensure that the exact observation specifications were replicated. However, we 
changed the Poisson seeds so that the noise added to the blank images was 
pulled from a unique distribution. We processed the blank images following 
the same \pipeline\ and custom steps as the images containing sources. For the 
$1/f$ noise removal, we masked out the features so they would not contribute 
to the measurement of the striping patterns. After flat-fielding the blank 
images, we normalized each image by its median and combined all 33 using a 
weighted mean. The weights were taken as $w_i =1/\mathrm{ERR}_i^2$, the 
uncertainty of the output flat field was estimated as $\sqrt{1/\sum{w_i}}$, 
and both the combined data and error arrays were normalized by the output 
image median. Figure~\ref{fig:flats} shows the six custom flats we created
for the SW detectors. We also show a custom flat for LW detector ALONG that 
we created for a previous version of our simulated images (SDR1). This ALONG 
feature is not present in our current simulations and is included here for
reference. 
We apply these six custom flats to the relevant detector images after the 
real flat is applied but before the flux calibration step that converts
the images to MJy/sr.

\subsection{Sky Subtraction} \label{sec:simsskysub}
The final stage of \pipeline\ includes a step (\texttt{SkyMatch}) that 
computes sky values in a collection of images in a way that matches the sky 
levels of several images before they are combined to form a mosaic. However, 
we found that the \texttt{SkyMatch} step does not properly remove the 
background in simulated CEERS images when run on a collection of images. This 
is likely due to a mismatch between the input photometric calibration 
parameters used by \mirage\ in simulating the data and those used by \pipeline.
Specifically, \mirage\ translates input magnitudes into count rates using 
\hst-style \texttt{PHOTFLAM} values derived from filter throughput curves. 
\mirage\ uses the same \texttt{PHOTFLAM} value for all short wavelength 
detectors for a given module and filter. The pipeline, however, converts count 
rates to MJy/sr using the \texttt{PHOTMJYSR} parameter in the flux calibration 
reference file, which depends on the pixel area and a mean gain value, both of 
which vary from detector to detector. A single value does not exist that can bring 
all simulated detector images to the same background level. Additionally, the 
CEERS dithers are not large enough to cover the gaps between detectors, and so 
there are many exposures with no overlap area in common for globally matching 
the sky values.

We find that the background levels in the final mosaics are significantly 
improved if \texttt{SkyMatch} is run on each calibrated image individually 
before mosaic creation. For this step, we adopt some minor changes to the 
default parameter values used to calculate the sky statistics. We tested
a grid of these parameter values and find that the following parameters yield output
images with median backgrounds closest to zero for the CEERS simulated data. 
We set the upper and lower sigma clipping limits to $2\sigma$ (from the 
default of $4\sigma$), the upper limit of usable pixel values for sky 
computation to 1 MJy/sr, and the number of clipping iterations to 10 
(default=5).

\subsection{Astrometry and Outlier Rejection} \label{sec:outliers}
Before creating mosaics, Stage 3 of \pipeline\ includes processing steps to 
apply astrometric corrections and identify outliers in the images. In real 
data, astrometric alignment is needed to account for uncertainties in 
guide star positions and correct any remaining distortions or offsets between 
images from different detectors.
However, as the simulated images are not created with astrometric errors or 
offsets from exposure to exposure, there is no need to perform this correction.

Next, the \texttt{OutlierDetection} step of \pipeline\ Stage 3 builds a stack of 
input images resampled onto a common grid in order to identify bad pixels or
cosmic rays that were not detected in the \texttt{Jump} step of Stage 1. 
We found that a bug in the \pipeline\ version 1.4.6 pipeline sometimes calculated 
incorrect WCS bounding boxes for the stacks. Occasionally, the bounding box  
would exclude a portion of some input images, resulting in all source flux 
from the region being flagged as outliers. Therefore, we use \pipeline\ version
1.5.2 for the \texttt{OutlierDetection} step. Once outlier pixels are
identified and flagged, we return to using version 1.4.6 for mosaicking for 
consistency with the rest of the reduction.

\begin{figure*}
\plotone{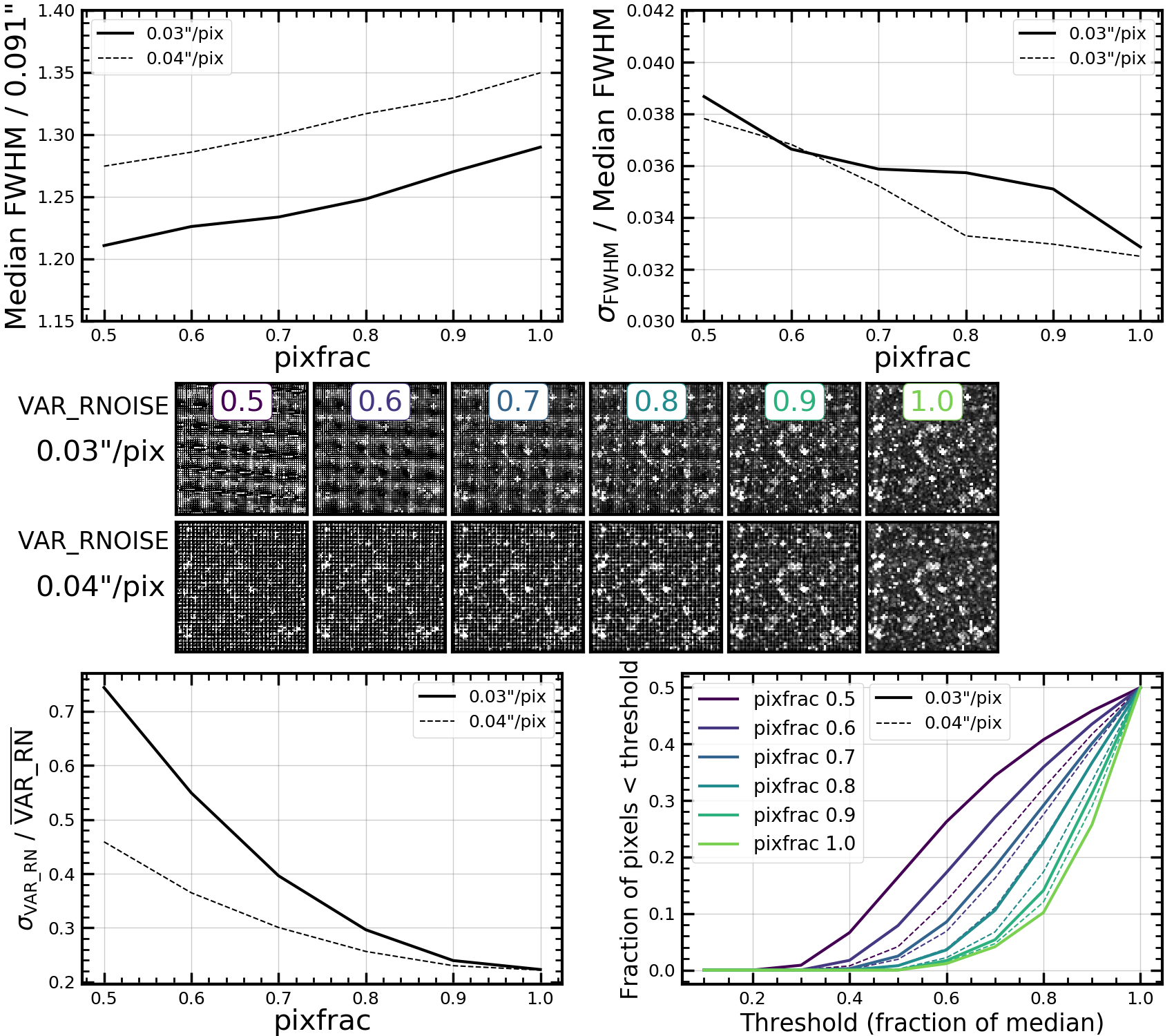}
\caption{Example of the drizzle tests we performed with simulated images containing 72 point sources and varying the output pixel scales and \texttt{pixfrac} values. The plots here show results for F277W, for which we explored 0\farcs03/pixel and 0\farcs04/pixel output pixel scales. In the top row we show the median recovered FWHM normalized by intrinsic FWHM (left) and the standard deviation of the 72 recovered FWHMs normalized by the median (right), both as a function of \texttt{pixfrac}. 
In the middle panels, we show small regions of the \texttt{VAR\_RNOISE} arrays for each pixel scale and \texttt{pixfrac}, all plotted with the same image stretch and scale. These panels illustrate the decreasing readnoise with increasing pixel scale and/or \texttt{pixfrac}. In the bottom left panel we show the 
normalized standard deviation of sigma-clipped pixel values in the 
\texttt{VAR\_RNOISE} as a function of \texttt{pixfrac}. In the bottom right panel we plot the fraction of pixels in the inverse \texttt{VAR\_RNOISE} arrays that lie below various thresholds as an way to quantify the number of outlier pixels in each mosaic.
These tests informed our choice to adopt an output pixel scale of 0\farcs03/pixel and a \texttt{pixfrac} of 1 for the mosaics in our first data release.
\label{fig:psfs}}
\end{figure*}

\begin{figure*}
\plotone{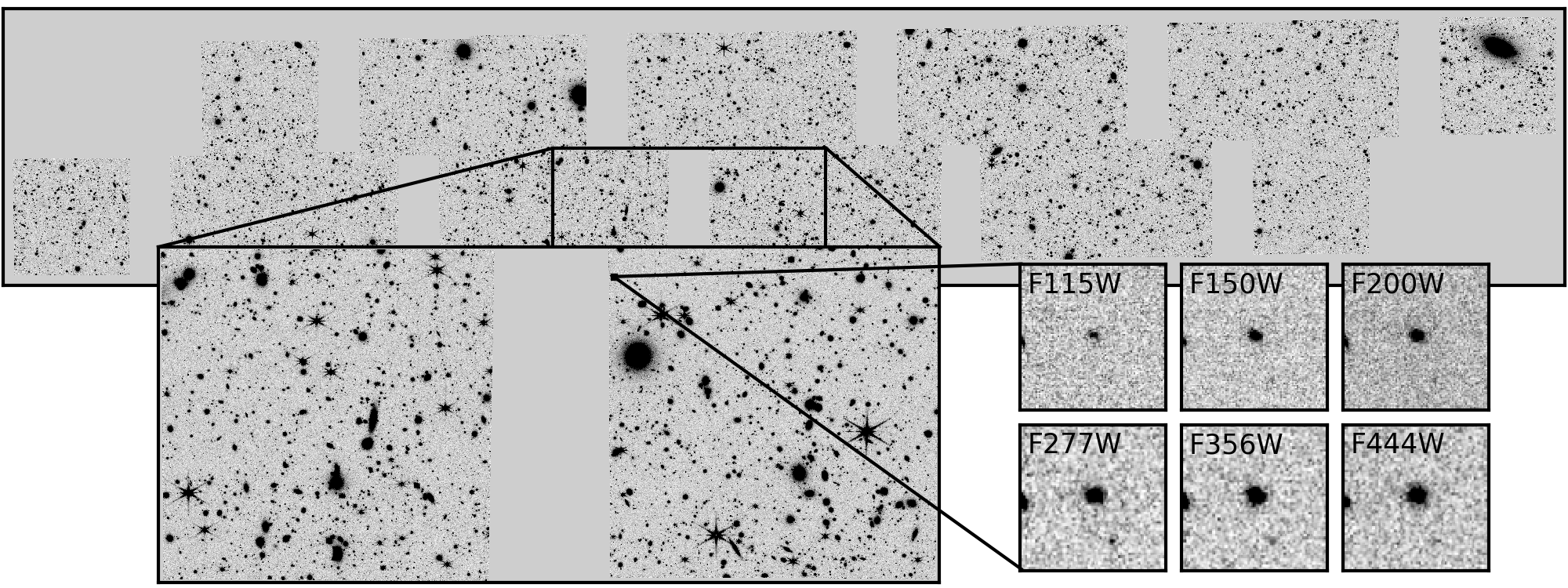}
\caption{The full CEERS simulated field, including all ten NIRCam pointings.
Here we show the F277W mosaic and a zoom-in around pointing 5, which was the 
focus of our early simulated data release. We further zoom in around one of 
the real $z\sim9$ candidate galaxies that has been added to this field
(see Section~\ref{sec:simsinputs}). We show 3\arcsec\ postage stamps of 
the simulated images of this galaxy in all filters. 
\label{fig:sims}}
\end{figure*}

\subsection{Simulated Mosaic Creation} \label{sec:simsmosaics}
Images are coadded in the Stage 3 step \texttt{Resample} as described in 
Section~\ref{sec:mosaics}, with an output pixel scale of 0\farcs03/pix
and \texttt{pixfrac}=1. We chose these drizzle parameters based on tests that we describe in the next paragraph, performed with simulated images of point sources. The mosaics are all drizzled onto a common 
WCS such that they are pixel-aligned across all filters, though they are 
not aligned with the \hst\ mosaics as the mock NIRCam sources do not have
counterparts in the \hst\ images. We note that due to pixels with a value of exactly zero in the variance arrays of input images, there are ``holes'' present in the error arrays of our simulated mosaics as described in Section~\ref{sec:mosaics}.

In order to determine the optimal output pixel scale and \texttt{pixfrac}, 
we created a simulated NIRCam pointing containing point sources and tested the
FWHM recovery for different output pixel scales and \texttt{pixfrac} values. 
We placed 72 point sources across the image, with a magnitude $m=23$ in all 
filters to avoid saturating the detectors in a CEERS-length exposure.
This PSF simulation uses the observation specification of a NIRCam imaging pointing obtained in parallel to the NIRSpec MSA primary observations, and so uses three small dithers determined by the NIRSpec nods. We explored output pixel scales of 0\farcs015/pixel and 0\farcs02/pixel in the SW images and 0\farcs03"/pixel and 0\farcs04"/pixel in the LW images. These choices are motivated by the need
for the pixel scales of the SW and LW images to be integer multiples of each other (and all NIRCam images to be integer multiples of the available \hst\ imaging)
to maintain alignment. We tested \texttt{pixfrac} values ranging from 0.5 to 1.

In Figure~\ref{fig:psfs}, we show a summary of the results of these tests for F277W as an example (comparing mosaics with with 0\farcs03/pixel and 0\farcs04/pixel). First, we measured the recovered FWHM of all 72 point sources. In the top left panel, we show the median recovered FWHM as a function of \texttt{pixfrac} normalized by the reported FWHM for this filter\footnote{The documented FWHMs are based on oversampled simulated PSFs created with WebbPSF. The PSFs will be updated during Cycle 1 based on in-flight measurements. See \url{jwst-docs.stsci.edu/jwst-near-infrared-camera/nircam-performance/nircam-point-spread-function}} (0.091\arcsec). 
The median of the recovered FWHMs increases towards larger \texttt{pixfrac}s, which is expected due to the increased pixel-to-pixel correlations in the drizzled images. The recovered FWHMs are $\sim$5\% larger for the 0\farcs04/pixel scale, indicating a preference for the smaller pixel scale. 
We also calculated the standard deviation of the recovered FWHMs, and plot these 
values normalized by the median recovered FWHM in the top right panel.
For both pixel scales, the normalized $\sigma_{\mathrm{FWHM}}$ decreases with 
increasing \texttt{pixfrac}, an expected trend as the output pixels more fully 
sample the PSFs.

Next we measured the noise properties in the drizzled mosaics. 
In the bottom left panel of Figure~\ref{fig:psfs}, we plot the standard deviation of the sigma-clipped pixel values in the \texttt{VAR\_RNOISE} maps normalized by the mean. These curves fall with increasing \texttt{pixfrac}, indicating that the width of the noise distribution decreases as more of each input pixel is incorporated into the output pixels. We also measured the fraction of pixels in the inverse variance maps 
(1/\texttt{VAR\_RNOISE}) that are a certain distance away from the median. In this 
way, we can quantify the number and magnitude of outlier pixels as a function of
\texttt{pixfrac}. These curves are shown in the lower right panel of Figure~\ref{fig:psfs}, color-coded by \texttt{pixfrac}. We once again see that the fraction of outliers decreases with increasing \texttt{pixfrac}. In the center of Figure~\ref{fig:psfs}, we show a small region of the \texttt{VAR\_RNOISE} maps for each output pixel scale and \texttt{pixfrac}. These postage stamps are all displayed with the same scale and stretch and provide a visual representation of how well the output pixels are sampled in each mosaic. The readnoise is reduced in mosaics with the larger pixel scale and/or larger \texttt{pixfrac}. 

These tests informed our decision to create mosaics on output pixel scales of 0\farcs03/pixel and \texttt{pixfrac}=1. This choice of \texttt{pixfrac} does result in maximum correlated noise introduced during the drizzle process, which we discuss in Section~\ref{sec:issues}. We note that a \texttt{pixfrac} of 0.9 or 0.8 may also be acceptable, and we will explore these options more completely in a future reduction. We plan to use these mosaics (both for the simulated and real NIRCam data) for source detection and photometry in all filters. For future data releases, we will create SW mosaics on an output scale of 0\farcs015/pixel for use characterizing source morphologies.

Finally, we note that the mosaic creation for the full CEERS simulated mosaic 
(all ten NIRCam pointings) requires a considerable amount of memory. This 
version of the pipeline holds all images, variance maps, and weight maps in 
memory while resampling each onto the common output grid. With 928 input 
images, the F115W mosaic is especially memory intensive, requiring almost one 
terabyte of memory to resample all input images onto a 0\farcs03/pixel output 
image (and an estimated 3.5 TB for an output 0\farcs015/pixel scale). We 
therefore use the Frontera computing system at TACC to create these mosaics.

\subsection{Simulated Raw and Reduced Data Products} \label{sec:sdrs}
We have provided two sets of simulated NIRCam images as part of public 
CEERS data releases. In CEERS Simulated Data Release 1 (SDR1\footnote{\url{ceers.github.io/sdr1.html}}), we shared a preliminary set of 
simulations of a single CEERS NIRCam pointing (labeled CEERS5), along with detailed step-by-step instructions for reducing the images in a Jupyter notebook. We also presented this work as part of JWebbinar 13. The mosaics released in SDR1 
had pixel scales of 0\farcs015/pixel in the SW filters, and 0\farcs03/pixel
in the LW filters. 
Next, we shared an updated simulation and reduction of this single NIRCam pointing in CEERS Simulated Data Release 3 (SDR3\footnote{\url{ceers.github.io/sdr3.html}}). This follow-up release incorporated improvements in our input mock galaxy catalog and \mirage, as well as updates to \pipeline\ and the reference files available at the time. These mosaics were all drizzled to an output pixel scale 
of 0\farcs03/pixel, and were pixel-aligned across all filters.
In Figure~\ref{fig:sims}, we show the full CEERS simulated mosaic, including
a panel zooming in to pointing 5, the pointing shared in SDR1 and SDR3. We further zoom in on simulated imaging in all
filters of one of the real $z\sim9$ candidates.

\bibliography{ceersnircam}

\end{document}